\documentclass[aps,prd,nofootinbib,onecolumn,superscriptaddress,preprintnumbers,balancelastpage,longbibliography]{revtex4-2}

\usepackage{amsmath}
\usepackage{slashed}
\usepackage{color, verbatim}
\usepackage{orcidlink}
\usepackage[normalem]{ulem}
\usepackage{bbold}
\usepackage[capitalise]{cleveref}
\usepackage{ragged2e}
\usepackage{caption}

\newcommand{\mr}[1]{{\color{blue}{#1}}}
\newcommand{\dd}[1]{{\color{orange}{#1}}}

\definecolor{mygreen}{rgb}{0.0, 0.5, 0.0}

\definecolor{royalpurple}{rgb}{0.47, 0.32, 0.66}

\newcommand{\pdv}[2]{\frac{\partial #1}{\partial #2}}
\newcommand{\pddv}[2]{\frac{\partial^2 #1}{\partial #2^2}}
\newcommand{\qq}[1]{\quad\text{#1}\quad}

\usepackage{xcolor}
\usepackage{amssymb}
\usepackage{graphicx}
\graphicspath{ {Images/} }

\begin{document}
\preprint{CERN-TH-2025-131}

\title{Resonant Landau-Zener Conversion In Multi-Axion Systems}

\author{David I. Dunsky\orcidlink{0000-0002-6323-8839}} 
\email{ddunsky@nyu.edu}
\affiliation{\small Center for Cosmology and Particle Physics, Department of Physics, New York University, New York, NY 10003, USA}

\author{Claudio Andrea Manzari\orcidlink{0000-0001-8114-3078}} 
\email{camanzari@berkeley.edu}
\affiliation{\small Leinweber Institute for Theoretical Physics, Department of Physics, University of California, Berkeley, CA 94720, USA}
\affiliation{\small Theoretical Physics Group, Lawrence Berkeley National Laboratory, Berkeley, CA 94720, USA}

\author{Pablo Qu\'ilez\orcidlink{0000-0002-4327-2706}}
\email{pquilez@ucsd.edu}
\affiliation{\small Department of Physics, University of California, San Diego, USA}

\author{Maria Ramos\orcidlink{0000-0001-7743-7364}}
\email{maria.ramos@cern.ch }
\affiliation{\small CERN, Theoretical Physics Department, Esplanade des Particules 1, Geneva 1211, Switzerland}

\author{Philip Sørensen\orcidlink{0000-0003-4780-9088}} 
\email{philip.soerensen@pd.infn.it}
\affiliation{\small Dipartimento di Fisica e Astronomia ‘G. Galilei’, Università di Padova, Via F. Marzolo 8, 35131 Padova, Italy}
\affiliation{\small INFN Sezione di Padova, Via F. Marzolo 8, 35131 Padova, Italy}

\begin{abstract}
Multiple axions may emerge in the low-energy effective theory of Nature. Generically, the potentials describing these axion fields are non-diagonal, leading to mass mixing between axion states which can be temperature-dependent due to QCD instanton effects.
As the temperature of the Universe drops, level crossing can occur, causing \textit{resonant} conversion between axion states.
In this work, we present an analytic study of the cosmological evolution of multi-axion systems including adiabatic and non-adiabatic resonant conversion from one axion state into another during the misalignment process. We show how the Landau-Zener formalism accurately captures the non-adiabatic resonant conversion, permitting an analytic description of the relic abundances of each axion field for nearly any arbitrary two-state axion mass matrix. As an application, we study the mixing of a QCD axion with an axion-like-particle for specific potentials to identify the predictions for haloscope experiments. We conclude that the detection of an axion off the expected QCD mass-coupling line predicts other haloscope targets if it mixes with the QCD axion.
\end{abstract}

\maketitle

\section{Introduction}

The Standard Model of particle physics (SM) has been remarkably successful in explaining a wide range of experimental observations across multiple sectors and energy scales. Despite this success, several experimental phenomena remain unexplained within the framework of the SM, highlighting the need for new physics. Notable examples include  dark matter, the origin of neutrino masses, and the hierarchical structure of flavor. Another important puzzle is the \textit{strong CP problem}, which arises from the lack of a theoretical explanation for the vanishing electric dipole moment of the neutron (nEDM). In the SM, the nEDM is induced by the only CP-violating operator in the Lagrangian of Quantum Chromodynamics (QCD),
\begin{align}
    \mathcal{L}_{\rm QCD} \supset \bar{\theta}\frac{g_s^2}{32\pi^2}\, G_a^{\mu\nu}\tilde{G}_{a,\, \mu\nu} \, ,
\end{align}
where $G_a^{\mu\nu}$ is the gluon field strength tensor, $g_s$ is the strong coupling constant, and $\bar{\theta} = \theta + {\rm arg\,det}(Y_u Y_d)$ is the physical -- \emph{i.e.} field redefinition invariant -- CP violating phase, which involves the quark Yukawa matrices $Y_u,\,Y_d$ . The parameter $\bar{\theta}$ is constrained by the experimental limit on the nEDM: $\bar{\theta} \lesssim 10^{-10}$~\cite{Pendlebury:2015lrz}, making it the smallest dimensionless parameter of the SM. A theoretical understanding of its smallness remains elusive.\\
A class of solutions to this problem is based on a classically exact, but anomalous chiral symmetry, which renders this parameter unphysical. Initially proposed by Peccei and Quinn~\cite{Peccei:1977hh,Peccei:1977ur}, the spontaneous breaking
of a global U(1) symmetry, anomalous with QCD, leads to a very light pseudoscalar field~\cite{Weinberg:1977ma,Wilczek:1977pj} that dynamically relaxes $\bar{\theta}$ to zero. This light pseudoscalar particle, known as the QCD \textit{axion}, is typically very weakly coupled to the SM and is predicted to be produced in the early Universe through the misalignment process \cite{Preskill:1982cy,Abbott:1982af,Dine:1982ah}, 
among other non-thermal production mechanisms \cite{Kibble:1976sj,Kibble:1980mv,Davis:1985pt,Davis:1986xc,Harari:1987ht,Battye:1993jv},
making it also an excellent dark matter candidate. Consequently, as a well-motivated extension of the SM, the QCD axion has seen in recent years both a growing theoretical interest and rapid progress in experimental searches. See \cite{OHare:2024nmr} for a recent review.

The QCD axion is an example of the more general class of axion-like-particles (ALPs), which are ultra-light weakly coupled pseudo-scalars appearing generically as pseudo-Goldstone bosons from spontaneously broken symmetries. Moreover, ALPs are now recognized as a generic feature of string theory compactifications~\cite{Svrcek:2006yi, Arvanitaki:2009fg,Cicoli:2012sz,Demirtas:2018akl,Halverson:2019cmy,Demirtas:2021gsq}, motivating the framework of the \textit{axiverse}, where the QCD axion is accompanied by a spectrum of axion-like particles. From now on we will generically refer to them as \textit{axions}. 

With the well-motivated scenario featuring both the QCD and other axions, it is interesting to ask how an additional axion might change the standard misalignment cosmology. Such axions produced in the early Universe generically mix quantum mechanically, and due to temperature-dependent masses generated by QCD, often exhibit level-crossings, or resonances, between the two states, which can significantly modify the standard relic misalignment abundances
\footnote{Other resonant conversions  include the Mikheyev–Smirnov–Wolfenstein (MSW) effect of solar neutrinos \cite{Wolfenstein:1977ue,Mikheyev:1985zog}, as well as in physics beyond the SM, such as conversion between photon and dark photons \cite{Mirizzi:2009iz}, photon and axions \cite{Mirizzi:2009nq}, neutrinos and sterile neutrinos \cite{Shi:1998km}, and the general framework of dark matter production via oscillations of Rapidly Oscillating Massive Particles \cite{ROMPs}.}.
The goal of this work is to study the misalignment process in general multi-axion models and provide analytic results for their relic abundances, including, for the first time, both adiabatic and non-adiabatic scenarios.

We now highlight the relationship of this paper to related works. Hill and Ross were the first to consider the cosmological production of the QCD axion and an ALP in a mixed two-state system  \cite{Hill:1988bu}, demonstrating how adiabatic level-crossing (resonance duration longer than the oscillation time between axion flavor states) 
can reduce the abundance of the QCD axion relative to its standard misalignment abundance \cite{Abbott:1982af,Dine:1982ah,Preskill:1982cy} for a given axion decay constant. Ref.\,\cite{Kitajima:2014xla} revisited the cosmological production of two axion systems, providing a more careful calculation of the relic abundances with analytic results for adiabatic level-crossings and numerical results for mildly non-adiabatic crossings. Similar works \cite{Ho:2018qur,Cyncynates:2023esj,Murai:2024nsp} studied the adiabatic level-crossing for a variety of coupled axion potentials, showing how the QCD axion misalignment abundance can be reduced or enhanced relative to the standard misalignment abundance depending on the structure of the mass matrix of the coupled axion system. The possibility of an analytic treatment of non-adiabatic level crossing (resonance width shorter than oscillation period) via Landau-Zener (LZ) was first mentioned in \cite{Muursepp:2024mbb}, though no analytic calculation was performed\,
\footnote{
While not a level-crossing, some coupled potentials can lead to a non-linear transfer of energy via phase-locking (autoresonance) when the two coupled ALPs have nearly identical and temperature-independent masses \cite{Cyncynates:2021xzw}.}.

In this work, we provide the first analytic treatment of non-adiabatic level crossings in two-state axion systems. In Sec.\,\ref{sec:convprob}, we review the features of the axion misalignment mechanism for a two-state axion system and compute the probability of conversion between axions in both the adiabatic and non-adiabatic regimes, using the LZ formalism. As an application, we apply the analytic LZ formula for the two-state axion potentials considered in \cite{Hill:1988bu, Kitajima:2014xla, Ho:2018qur,Cyncynates:2023esj,Murai:2024nsp,Muursepp:2024mbb,Gavela:2023tzu} showing strong concordance with numerical cross-checks. We also highlight the \textit{maxion} scenario, introduced in Ref.~\cite{Gavela:2023tzu}, in which the role of QCD axion is maximally split across multiple mass eigenstates. In Sec.\,\ref{sec:DMabundance} we use these results to compute the relic densities of the two axions and in Sec.\,\ref{sec:conclusions}, discuss the implications for various axion-bounds. A rigorous derivation of the LZ formula for a coupled axion-system is discussed in Appendix \ref{app:LZ}.

\section{Probability of Conversion: Adiabatic and Non-Adiabatic Regimes}\label{sec:convprob}
The equation of motion of a two-state axion system, $A\equiv(a, a_S)^T$,
possessing a potential $V(a,a_S)$, 
is 
\begin{align}
    (\partial_t^2 - \nabla^2 + 3H \partial_t) \,
    \begin{pmatrix}
        a
        \\
        a_S
    \end{pmatrix}
    =-\mathbf{M}^2 
    \begin{pmatrix}
        a
        \\
        a_S
    \end{pmatrix}\, ,
    \label{eq:AxionAlpEOM}
\end{align}
where $a$ and $a_S$ are the two axion flavor states, and $H = \dot{R}(t)/R(t)$ is the Hubble parameter in a FLRW Universe with scale factor $R(t)$. On the right-hand-side of Eq.\,\eqref{eq:AxionAlpEOM}, we expand the axion potential to quadratic order to yield a squared mass matrix, $\mathbf{M^2}$, which in general, is non-diagonal with elements
\begin{align}
    \mathbf{M}^2 
    &= 
    \begin{pmatrix}
        m_{aa}^2 && m_{as}^2
        \\[2pt]
        m_{as}^2 && m_{ss}^2
    \end{pmatrix}
    \label{eq:exMassSq}
    \, .
\end{align}
As is common in the study of damped harmonic oscillators, one can rescale the axion states to eliminate the friction term, $3H\partial_t A$.
In this case, it corresponds to defining the comoving axion field, $\tilde A=R(t)^{3/2}A$. The transformed equation of motion reads,
\begin{align}
       \Big(\partial_t^2 +\frac{3}{4}H^2+\mathbf{M}^2\Big) \tilde A=0\,,
    \label{eq:AxionAlpEOM2}
\end{align}
where we neglect the gradient term by assuming spatially homogeneous axion fields, a good approximation for pre-inflationary axions, or equivalently, when the reheat temperature of the Universe is below the $U(1)$ symmetry breaking scale associated with each axion. 

Eq.\,\eqref{eq:AxionAlpEOM2} is a system of two coupled oscillators with no friction and a time-dependent frequency matrix $\boldsymbol{\omega}^2(t)\equiv \frac{3}{4}H^2\mathbb{1}+\mathbf{M}^2$, where the identity matrix has been dropped from Eq.~\eqref{eq:AxionAlpEOM2} for convenience.
The dynamics of this system are most simply understood by subsequently transforming 
to the \textit{instantaneous} mass basis of the axion fields, $(a_H, a_L)$, which is the basis that diagonalizes $\mathbf{M}^2$ at a given point in time,
\begin{align}
  & \begin{pmatrix}
        m^2_{H} && 0
        \\
        0 && m^2_{L}
    \end{pmatrix}\equiv \boldsymbol{M}_D^2= \boldsymbol{U}^\dagger \boldsymbol{M}^2 \boldsymbol{U}; &\begin{pmatrix}
        a
        \\
        a_S
    \end{pmatrix}
    =
    \boldsymbol{U}
    \begin{pmatrix}
        a_H
        \\
        a_L
    \end{pmatrix}
    =
    \begin{pmatrix}
        \cos \xi && -\sin \xi
        \\
        \sin \xi && \cos \xi
    \end{pmatrix}
    \begin{pmatrix}
        a_H
        \\
        a_L
    \end{pmatrix}
    \, .
    \label{eq:rot}
\end{align}
Note that $\boldsymbol{U}(t)$, and hence the mixing angle $\xi(t)$, is time-dependent because some of the matrix elements of $\mathbf{M}^2$ will inherit the temperature dependence of $SU(3)_c$ instanton effects which become important at temperatures near the QCD phase transition. 
Diagonalizing $\mathbf{M}^2$ gives the mass eigenvalues $m_{H,L}$
 and mixing angle $\xi$,
\begin{align}
   & m_{H,L}^2(t) = \frac{1}{2}\big[ m_{aa}^2(t) + m_{ss}^2(t) \pm 
    \sqrt{(m_{aa}^2(t) - m_{ss}^2(t))^2 + 4 (m_{as}^2(t))^2}\, \big]\,;
&  \tan 2 \xi(t) 
    &= \frac{2 m_{as}^2(t)}{m_{aa}^2(t) - m_{ss}^2(t)}
    \, ,
    \label{eq:tan2xi}
\end{align}
where we explicitly indicate the time-dependence to emphasize that $m_H^2$, $m_L^2$, and $\xi$ are functions of temperature and hence time. 

\begin{figure}[t]
\includegraphics[width=0.49\textwidth]{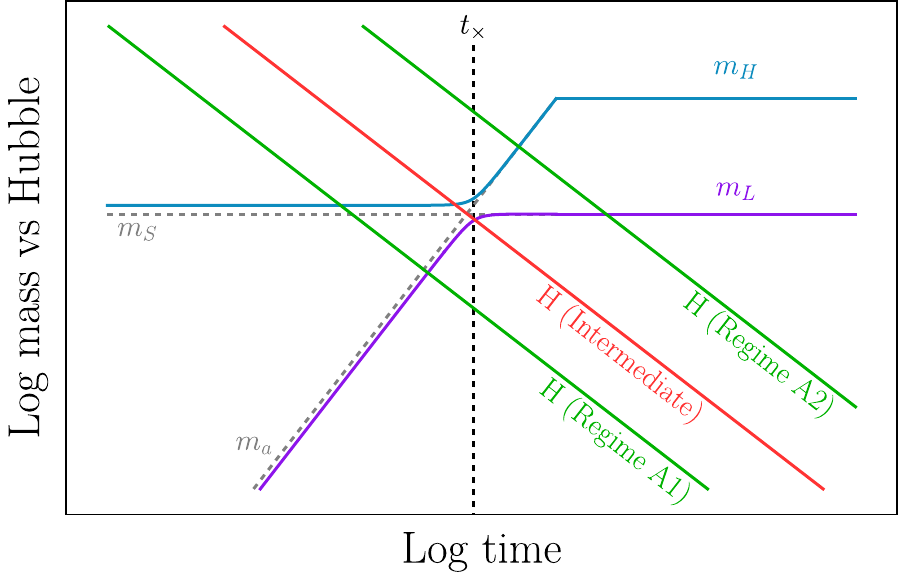}
\includegraphics[width=0.49\textwidth]{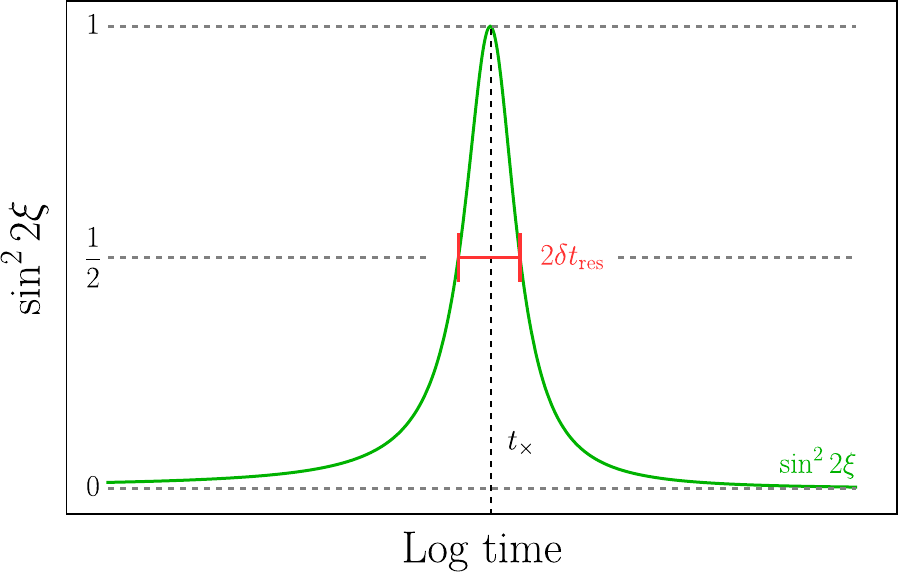}
\caption{\justifying\textbf{Left:} Sketch of the avoided crossing that underlies the phenomenology in this paper. The axion fields start oscillating when $m_{H,L}\sim H$. This results in different scenarios depending on whether the onset of oscillations take place before or after the crossing. The physical mass eigenvalues $m_H$ and $m_L$ are indicated by colored lines. Several potential scenarios for the Hubble parameter $H$ are shown and colored 
to match the different regimes in Fig. \ref{fig:comparison}. The masses of the flavor states $m_a(T)$ and $m_S$, using the conventions of Eq.\,\eqref{eq:VTAK}, are represented by the gray dashed lines. 
\textbf{Right:} Sketch of the evolution of the mixing parameter $\sin^2 2\xi$ from which we define the width of the resonance, $\delta t_{\rm res}$. Note that the right sketch shows a much shorter time interval than the left sketch so that both can highlight the relevant features.} 
\label{fig:masscross}
\end{figure}

Coupled axion systems -- much like other coupled oscillators such as diatomic molecules or neutrinos -- can exhibit an interesting phenomenon: \textit{avoided level crossing}~\cite{wigner1929uber}. This effect occurs when the two axion mass~eigenvalues approach degeneracy -- and would be degenerate but  for their off-diagonal interaction $m_{as}^2$, which prevents an exact crossing -- as shown qualitatively in \cref{fig:masscross} (left panel).
At this (avoided) level crossing, the two states are maximally mixed, with the rotation angle, $\xi$, in $\mathbf{U}$ going to $\pi/4$, or equivalently,  $|\tan 2\xi| \rightarrow \infty$. 

For axions, a level-crossing occurs if at high temperatures in the early Universe, $m_{ss}^2 > m_{aa}^2$ (or vice versa), but at low temperatures in the late Universe, $m_{aa}^2 > m_{ss}^2$ (or vice versa). At some instant in time, $t_\times$, it must be then that $m_{aa}(t_\times)^2 = m_{ss}(t_\times)^2$, which from Eq.\,\eqref{eq:tan2xi} implies $|\tan 2\xi| \rightarrow \infty$. In this cosmology, the value of $\sin^2 2\xi$ traces out a resonance-like curve in time, sharply peaking at $t_\times$ when it briefly becomes unity, as depicted in Fig.~\ref{fig:masscross} (right panel).  Indeed, this \textit{resonance} structure of $\sin^2 2\xi$ generates efficient probability of conversion between $a$ and $a_S$ states, and depending on how sharply peaked (non-adiabatic) the resonance is, strong conversion between $a_H$ and $a_L$ too. For this reason, we will interchangeably call $t_\times$ the crossing time or resonance time, as defined by
\begin{align}
    \label{eq:tCross}
    m_{aa}^2(t_\times) - m_{ss}^2(t_\times) = 0 \quad (\text{Definition of the 
    crossing time $t_\times$)} \, .
\end{align}

The equation of motion, Eq.\,\eqref{eq:AxionAlpEOM2}, in terms of the instantaneous comoving mass eigenstates $(\tilde{a}_H \;\; \tilde{a}_L)^T = \mathbf{U}^\dagger(t) (\tilde{a} \; \; \tilde{a}_S)^T$ is
\begin{equation}
    \partial_t^2 \,
    \begin{pmatrix}
        \tilde{a}_H
        \\
        \tilde{a}_L
    \end{pmatrix}
    = 
    \begin{pmatrix}
        \dot{\xi}^2-\frac{3}{4}H^2-m_H^2
        && ( \ddot{\xi}+2\dot{\xi}\partial_t)
        \\[2pt]
        -( \ddot{\xi}+2\dot{\xi}\partial_t) && 
        \dot{\xi}^2-\frac{3}{4}H^2-m_L^2
    \end{pmatrix}
    \begin{pmatrix}
        \tilde{a}_H
        \\
        \tilde{a}_L
    \end{pmatrix}\, .
    \label{eq:AxionAlpEOMMassBasisSimplified}
\end{equation}
The behavior of this system of coupled differential equations is governed by the hierarchy between the variation of the mixing angle $\dot \xi$ and the oscillation rates of each individual axion, $m_{H,L}$, and the oscillation rate \textit{between} axions, $m_H - m_L$, the non-relativistic analog of a neutrino oscillation rate. Depending on this interplay, two distinct regimes emerge: the adiabatic and non-adiabatic limits, corresponding to whether $\dot{\xi}$ can be neglected or not, which is tied to the existence of a resonance.

\paragraph*{\textbf{Adiabatic case.}}

When the mixing angle changes slowly enough in time that the $\dot \xi$ and $\ddot \xi$ terms in Eq. \eqref{eq:AxionAlpEOMMassBasisSimplified} can be neglected,  the equation of motion for the comoving mass eigenstates $\tilde{a}_H$ and $\tilde{a}_L$ is nearly diagonal,
\begin{equation}
    \partial_t^2 \,
    \begin{pmatrix}
        \tilde{a}_H
        \\
        \tilde{a}_L
    \end{pmatrix}
    \simeq 
    \begin{pmatrix}
        -\frac{3}{4}H^2-m_H^2
        && 0
        \\[2pt]
        0 && 
        -\frac{3}{4}H^2-m_L^2
    \end{pmatrix}
    \begin{pmatrix}
        \tilde{a}_H
        \\
        \tilde{a}_L
    \end{pmatrix}\, .
    \label{eq:AxionAlpEOMMassBasisAdiabatic}
\end{equation}
In this regime, the \textit{adiabatic theorem} implies that this slow-acting perturbation in the axion Hamiltonian does not affect the energy level populations. As a consequence, states initially occupying an energy eigenstate will remain within that energy eigenstate, as they evolve continuously in flavor space. More concretely, after the fields start to oscillate at time $t_{{\rm osc},H,L}$ when $m_{H,L} > H$, one can solve the independent evolution of the two mass eigenstates $a_H$ and $a_L$ under the WKB approximation,
\begin{align}
    & a_H(t) = a_{H,\text{osc}}\sqrt{\frac{{R_{\text{osc}}^3 m_{H,\text{osc}}}}{{R^3 m_H}}}  \cos\left(\int^t_{t_{\text{osc,}H}} m_H(\tau) \mathrm{d}\tau\right) \, ,
    & a_L(t) = a_{L,\text{osc}}\sqrt\frac{{R_{\text{osc}}^3 m_{L,\text{osc}}}}{R^3 m_L}  \cos\left(\int^t_{t_{\text{osc,}L}} m_L(\tau) \mathrm{d}\tau\right) \, ,
    \label{Eq:WKBexactsolmain}
\end{align}
where  the `\rm{osc}' subscript indicates that the associated quantity is evaluated at the $t = t_{{\rm osc},H}$ for the heavy field and $t_{{\rm osc},L}$ for the light field.
The WKB approximation is equivalent to the assumption that the mass eigenvalues 
change slowly,
$\dot\omega_i/\omega_i^2\ll1$ where $\omega_i^2=3/4H^2+m_i^2$ for $i = \{H,L\}$, implying $H\ll m_i$ and $\dot m_i/m_i^2\ll1$ . 
Thus, even if the mass eigenstates change in time, there exists two independent  adiabatic invariants i.e. comoving number of axions, $n_H^{\rm WKB}$ and $n_L^{\rm WKB}$, which remain constant in the evolution,
\begin{align}
    &n_H^{\rm WKB}(T) = \frac{\rho_H(T) R^3}{m_H(T)} \simeq \text{constant} \, ,
    &n_L^{\rm WKB}(T) = \frac{\rho_L(T)R^3}{m_L(T)} \simeq \text{constant} \label{eq:numberdensitiesWKB} \, ,
\end{align}
where $\rho_{H,L}\equiv \dot a^2_{H,L}/2+ m_{H,L}^2 a^2_{H,L}/2 $ are the corresponding energy densities of the two mass eigenstates.
It is important to stress that in terms of any static basis, e.g. the flavor basis $a$ and $a_S$, the adiabatic evolution of the fields is  coupled and it is only in the instantaneous eigenstate basis that the two evolutions decouple.
Note that the WKB results, \cref{Eq:WKBexactsolmain,eq:numberdensitiesWKB}, apply even when the axion fields experience a resonance (see Fig.\,\ref{fig:masscross} and resonance condition\, \eqref{eq:tCross}), as long as the evolution remains adiabatic ($\dot\xi\ll\Delta m$, see \cref{eq:gamma}), analogously to the Mikheyev–Smirnov–Wolfenstein (MSW) effect in solar neutrinos \cite{Mikheyev:1985zog,Wolfenstein:1977ue}.
We now quantify the condition for non-adiabatic evolution and show how such evolution is captured with the LZ formalism.
\paragraph*{\textbf{Non-adiabatic case.}}
What happens when $\dot{\xi}$ is large so that the off-diagonal elements of \eqref{eq:AxionAlpEOMMassBasisSimplified} become important in the evolution of $(a_H, a_L)$? In this \textit{non-adiabatic} regime, the population of the mass eigenstates $a_H$ and $a_L$ can convert between themselves. To understand any non-adiabatic evolution, it is useful to linearize Eq.\,\eqref{eq:AxionAlpEOMMassBasisSimplified} to a two-state Schrödinger equation, which we derive for the axion system in Appendix \ref{app:FeshbachVillars}. The resulting equations of motions, written in terms of positive and negative frequency modes of the axion field, are

\begin{align}
    \label{eq:schrodingerEOM}
    i \partial_t
    \begin{pmatrix}
        \tilde{a}_{H \pm}
        \\
        \tilde{a}_{L \pm}
    \end{pmatrix}
    &= 
    \pm
    \begin{pmatrix}
        \frac{m_H + m_L}{2} && 0
        \\
        0 &&  \frac{m_H + m_L}{2}
    \end{pmatrix}
    \begin{pmatrix}
        \tilde{a}_{H \pm}
        \\
        \tilde{a}_{L \pm}
    \end{pmatrix}
    +
    \begin{pmatrix}
        \pm\frac{m_H - m_L}{2} - \frac{i}{2}\frac{\dot{m_H}}{m_H}  &&  i \dot{\xi}
        \\
        -i \dot{\xi} &&  \mp\frac{m_H - m_L}{2}  - \frac{i}{2}\frac{\dot{m_L}}{m_L} 
    \end{pmatrix}
    \begin{pmatrix}
        \tilde{a}_{H \pm}
        \\
        \tilde{a}_{L \pm}
    \end{pmatrix} \, ,
\end{align} 
where $\tilde{a}_{H} \equiv \tilde{a}_{H+} +  \tilde{a}_{H-}$ and  $\tilde{a}_{L} \equiv \tilde{a}_{L+} +  \tilde{a}_{L-}$. The derivation of Eq.\,\eqref{eq:schrodingerEOM} assumes the axion fields have negligible momenta so that any spatial gradients are small and that $m_{H,L} > H$, which allows the axion fields to oscillate. 

Since the term proportional to the identity matrix gives rise to an unimportant phase, Eq.~\ref{eq:schrodingerEOM} makes manifest that non-adiabatic evolution occurs when $\dot{\xi} \gg \omega_{\rm osc}$,
where
\begin{align}
    \omega_{\rm osc} = E_H - E_L = \Delta m = \frac{m_H^2 - m_L^2}{m_H + m_L}
    \label{eq:OscFrequencyDifference}
\end{align}
is the oscillation rate \textit{between} the two axion flavor states, analogous to the oscillation rate between neutrino flavors. The larger $\dot{\xi}$ is relative to $\omega_{\rm osc}$, the greater the probability of conversion between $\tilde{a}_H$  and $\tilde{a}_L$ since $\dot{\xi}$ connects $\tilde{a}_H$ and $\tilde{a}_L$ states in the Hamiltonian of Eq.\,\eqref{eq:schrodingerEOM}. In terms of the rate of change of the matrix elements of $\mathbf{M}^2$, $\dot{\xi}$ can be found by taking the time derivative of Eq. \eqref{eq:tan2xi},
\begin{align}
\begin{split}
      \dot \xi &= \frac{1}{4 m_{as}^2}\left[ 2\sin 2\xi \cos 2 \xi  \frac{d m_{as}^2}{dt} - \sin^2 2\xi \left(\frac{d m_{ss}^2}{dt} - \frac{d m_{aa}^2}{dt} \right)\right] \, .
\end{split}
\label{eq:2xiDot}
\end{align}
Eq.\,\eqref{eq:2xiDot} demonstrates that the rapid changes in the mixing angle $\xi$ occur prominently at $t_{\rm \times}$  when $\sin^2 2\xi$ briefly equals unity.
The duration of this resonance, $\delta t_{\rm res}$, is set by the width of the $\sin^2 2\xi$ peak as shown in the right panel of Fig.\,\ref{fig:masscross}. It is convenient to define $\delta t_{\rm res}$  by the condition that if $\sin^2 2\xi(t_\times) = 1$, then $\sin^2 2\xi(t_{\times} + \delta t_{\rm res}) = 1/2$ \cite{kuo1989nonadiabatic}. From Eq.\,\eqref{eq:tan2xi}, this gives 
\begin{align}
\label{eq:tres}
    \delta t_{\rm res} = 2\left|\frac{ m_{as}^2}{\frac{d m_{ss}^2}{dt} - \frac{d m_{aa}^2}{dt}}\right|_{t = t_\times} = \frac{1}{2|\dot{\xi}(t_\times)|}\,.
\end{align}
Thus, the degree of non-adiabaticity is tied to the existence of a resonance.  

The evolution of the $(\tilde{a}_H, \tilde{a}_L)$ system for all time, including the brief non-adiabatic behavior around $t_{\times}$, can be analytically determined by solving Eq.~\ref{eq:schrodingerEOM} along a world-line that is $\textit{always}$ adiabatic, where the evolution of the $(\tilde{a}_H, \tilde{a}_L)$ states are known and is simply the WKB evolution. This method, due to Landau \cite{Landau:1932vnv,landau1977quantum}, is achived by analytically continuing the worldline into the complex plane around $t_{\times}$ such that non-adiabatic evolution is avoided as shown in App~\ref{app:Landau_Approach}. Alternatively, Eq.\eqref{eq:schrodingerEOM} can be solved directly in terms of parabolic Cylindrical functions, which is the method of Zener \cite{zener1932non}.
\footnote{
Interestingly, in addition to the derivations by Landau and Zener, analogous results were obtained in the same year (1932) by Majorana~\cite{Majorana:1932ga} and St\"uckelberg~\cite{stueckelberg1932}, in the contexts of oriented atoms in a varying magnetic field and inelastic atomic collisions, respectively.}
We confirm both methods give the same result in Appendix \ref{app:Zener_Approach}.

The key result is that, up to a phase, 
each axion field after $t_\times$ becomes a linear combination of $\tilde{a}_H$ and $\tilde{a}_L$ given by the solution \eqref{eq:adiabaticSolution}, %
\begin{align}
    {a}_H(t_\times^+) = C_1 \,{a}_H(t_\times^-) + C_2 {a}_L(t_\times^-) \, ,
\end{align}
with the LZ probability of conversion, $P_{\rm LZ}$, given by
\begin{align}
    \label{eq:LZProbability}
   |C_2|^2 = P_{\rm LZ}(\tilde{a}_H \leftrightarrow \tilde{a}_L) = \exp(- \pi \gamma/2)\,\quad \text{and}\quad |C_1|^2 = 1 - |C_2|^2\, ,
\end{align}
where the \textit{adiabatic parameter} $\gamma$ given by
\begin{align}
    \gamma  = \left\vert \frac{4(m_{as}^2)^2}{\frac{d \, m_{aa}^2}{dt} - \frac{d\, m_{ss}^2}{dt}}\frac{1}{m_H + m_L}  \right\vert _{t=t_{\times}}  = \Big|\frac{\omega_{\rm osc}}{2\dot\xi}\Big|_{t=t_\times}= \delta t_{\rm res}\omega_{\rm osc}(t_\times) \, .
    \label{eq:gamma}
\end{align}
In the second equality, we have made use of the fact that $m_H^2 - m_L^2 = 2 |m_{as}^2|$  at $t_{\times}$ (see Eq.\,\eqref{eq:tan2xi}), to write $\gamma$ in terms of $\omega_{\rm osc}(t_\times)$ and $\delta t_{\rm res}$.~We point out that, while the parametric dependence of $\gamma$ on the oscillation rate between the two axions has been suggested in previous studies~\cite{Ho:2018qur, Murai:2024nsp}, the analytical expression in Eq.~\ref{eq:LZProbability} is derived for the first time in this work.

Whenever $\gamma \gg 1$, the evolution is adiabatic through crossing, which corresponds to $P_{\rm LZ} \to 0$, and the result reduces to~\cref{Eq:WKBexactsolmain}. In the opposite limit, $\gamma \ll 1$, the crossing becomes highly non-adiabatic, and the LZ probability captures the likelihood of conversion between mass eigenstates, with $P_{\rm LZ} \to 1 - \pi\gamma/2\sim 1$. In this regime, the populations of the mass eigenstates efficiently exchange.

Two important assumptions underline the LZ treatment:
during the level crossing, \textit{(1)} the difference between the 
diagonal elements in $\mathbf M^2$ evolves linearly with time and \textit{(2)} the variation of the off-diagonal element is sufficiently small in comparison to the latter, so that it can be neglected. As we show in App~\ref{app:LZValidity}, \textit{such conditions are only violated if the resonance width is larger than the age of the Universe}, which does not occur unless the two derivatives, $dm_{aa}^2/dt$ and $dm_{ss}^2/dt$ in the expression above are tuned to cancel each other at $t_{\times}$. We can therefore employ the LZ method in the entire non-adiabatic regions for most axion two-state systems. 

In the following sections, we study two different axion potentials and discuss the mixing in different regimes. We will analytically determine the relic abundance of the two mass eigenstates today, $\Omega_{a_L}$ and $\Omega_{a_H}$, in the adiabatic regime and the non-adiabatic regime. Furthermore, we will also solve the coupled equations of motion numerically to confirm the analytic results.

\subsection{Application I}\label{sec:PotentialI}
Let us start by considering the following interaction potential~\cite{Murai:2023xjn}: 
\begin{equation}
V(a,\,a_S)  =  m_a^2(T) f_a^2 \left[1-\cos\left(\frac{a}{f_a}\right)\right]  +m_S^2f_S^2\left[1-\cos\left(\frac{a_S}{f_S}+\frac{a}{f_a}\right)\right]\,,
\label{eq:VTAK}
\end{equation}
where the first term stems from the interaction with gluons, $\frac{\alpha_s}{8\pi} \frac{a}{f_a} G \widetilde G$, being the overall amplitude the QCD  topological susceptibility, $ \chi (T) = m_a^2(T) f_a^2  $.

By expanding the potential to second order in the scalar fields, one identifies the mass matrix:
\begin{equation}
V (a,\,a_s)  \approx
\begin{pmatrix} 
a_S & a 
\end{pmatrix} 
\left(\begin{array}{cc}
m_S^2 & m_S^2 R_f \\
m_S^2 R_f & m_a^2(T)+ m_S^2 R_f^2
\end{array}\right) \begin{pmatrix} a_S \\  a \end{pmatrix} \equiv A^T \mathbf{M}^2 A\,,
\label{eq:massMatrix}
\end{equation}
where $R_f \equiv f_S/f_a$ and $R_m \equiv m_S/m_{a,0}$. 
The temperature dependence of the QCD topological susceptibility follows a power-law at high temperatures and remains constant below $T_{\rm QCD}$,
\begin{equation}
m_a^2 (T) = m_{a,0}^2 \text{max} \left\{1, \left(\frac{T}{T_{\rm QCD}}\right)^{-2n}\right\}\,, \label{eq:maofT}
\end{equation}
with $m_{a,0}^2 \equiv m_a^2(T=0)= \frac{m_\pi^2 f_\pi^2}{f_a^2} \frac{{m_u\,m_d}}{(m_u+m_d)^2}$, $T_{\rm QCD} \approx 100$ MeV and $n\approx { 3.34}$\,\footnote{ Note that there are several estimates for $n$. While the power-law behaviour predicted by the DIGA for QCD, $n\sim 4$, agrees with some lattice simulations \cite{Borsanyi:2016ksw}, alternative approximations like the interacting instanton liquid model (IILM) suggest $n \sim 3.3$ and other lattice computations result in smaller values, $n \sim 3.8-4.2$ \cite{Trunin:2015yda} for QCD.}.

In this scenario, the eigenvalues read
\begin{equation}
    m_{H,L}^2=\frac{1}{2} m_a^2 + m_s^ 2 \left(1+R_f^2\right) \pm \sqrt{m_a^2 - 2 m_a^2 m_s^ 2 (1-R_f^2) + m_s^4 (1+ R_f^2)^2}\,,
\end{equation}
whereas the mixing angle is given by
\begin{equation}
\tan 2\xi = \frac{2 R_f}{1-R_f^2 - (1/R_m^2) (m_a^2/m_{a,0}^2)}\,.
\label{eq:angle_pot1}
\end{equation}

Focusing on the high temperature limit of Eq.\,\eqref{eq:VTAK}, one can also obtain the initial field values of the two mass eigenstates as a function of the initial misalignment angles $\theta_{a}\equiv a/f_a$ 
 and $\theta_{s}\equiv a_s/f_s$:
\begin{align}
   & a_{H} (T)= \frac{R_f f_a}{\sqrt{1+R_f^2}}(\theta_{a}+\theta_{s})\,,\quad 
    &a_{L} (T)= \frac{R_f f_a}{\sqrt{1+R_f^2}}\Big(\frac{1}{R_f}\theta_{a}-R_f\theta_{s}\Big)\,.
    \label{Eq:InitialFieldValues}
\end{align}
These expressions will be relevant to the discussion that follows.

\begin{figure}[t]
\centering
\includegraphics[width=0.46\textwidth]{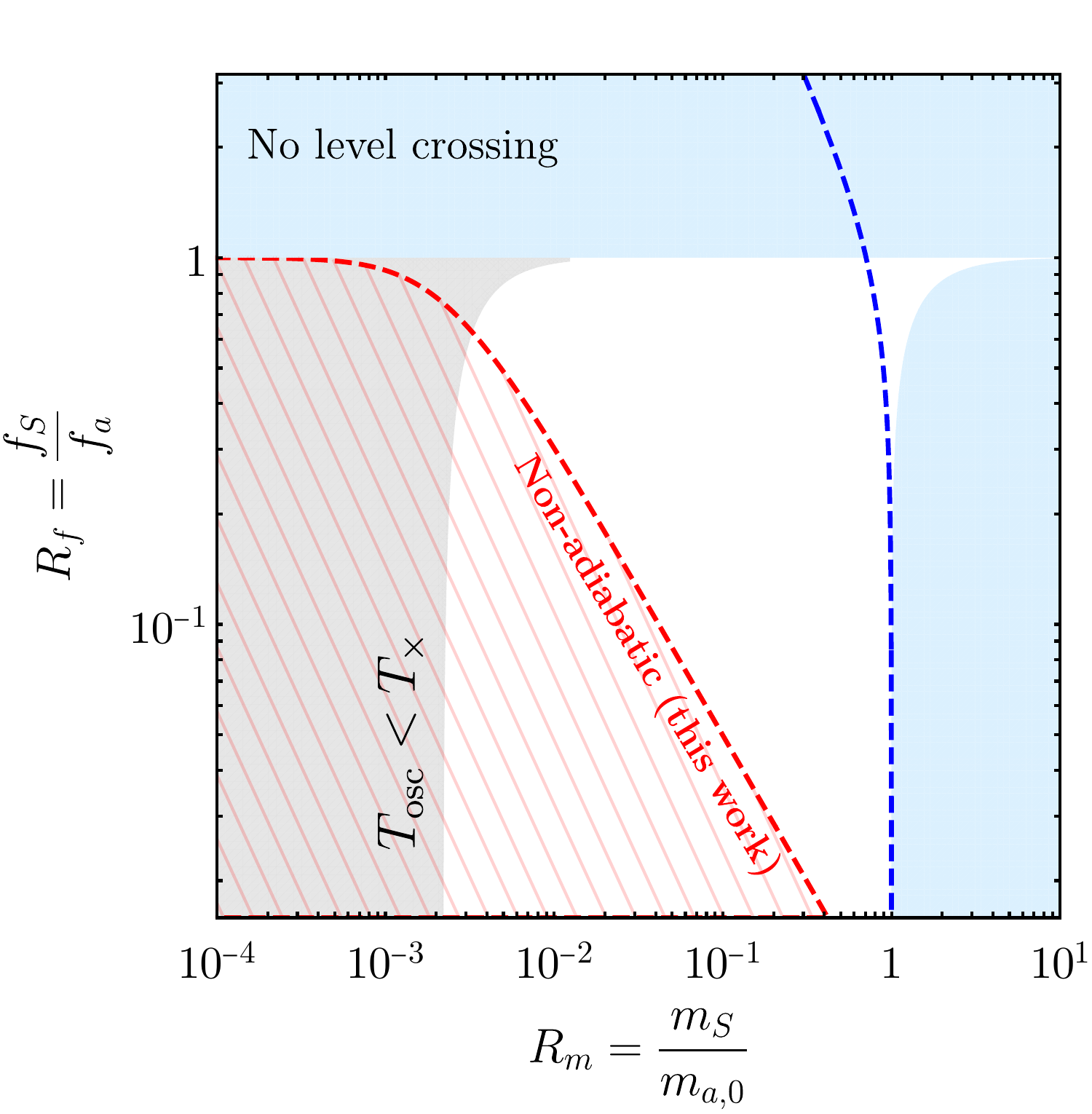}
\caption{\justifying Regions where level crossing can take place, assuming the potential in Eq.~\ref{eq:VTAK} with $f_a = 10^{13}$\,GeV. The dashed red line corresponds to the adiabaticity parameter $\gamma=1$. 
Our work presents a novel analytical procedure to obtain the dark matter densities in the non-adiabatic ($\gamma < 1$) regime, represented by the red dashed region in the figure.
 The blue dashed line highlights the relation between $R_f$ and $R_m$ imposed by the QCD maxion model; see Sec.~\ref{sec:haloscope} for details.
}
\label{fig:regionLC}
\end{figure}

As argued in the last section, the temperature at which level crossing takes place, $T_\times$, can be found by exploring the resonant behavior of the mixing angle, defined when $|\tan 2 \xi(T_\times)| \rightarrow \infty$\,: 
\begin{align}
 \frac{m_a^2 (T_{\times})}{m_{a,0}^2} = R_m^2 (1-R_f^2)\,
\Leftrightarrow T_{\times} = \left({\sqrt{1-R_f^2} R_m}\right)^{-1/n} T_{\rm QCD}\,.
\label{eq:Tcross}
\end{align}
The first equation above has solutions whenever $R_f < 1$ and  $R_m^2 (1-R_f^2)<1$, 
therefore level crossing is forbidden in the blue regions represented in Fig.~\ref{fig:regionLC}. These regions have been previously identified in Ref.~\cite{Ho:2018qur}.


With these results, it is now straightforward to obtain the LZ probability in terms of the adiabatic parameter defined in Eq.~\ref{eq:gamma}, which in this model reads
\begin{align}
\gamma 
&  = \frac{m_S}{H(T_\times) n} R_f^2 + \mathcal{O}(R_f^4)\,.
\label{eq:gammaTAK}
\end{align}
In Fig.~\ref{fig:regionLC}, we represent the parameter space where $\gamma < 1$ enclosed by a dashed red line, assuming that $f_a = 10^{13}$\,GeV. By employing the LZ conversion probability, we will show that the evolution of the two fields can be studied analytically in this entire region, as long as the fields start oscillating~{before} $T_\times$.

\subsection{Application II}\label{sec:PotentialII}

To further illustrate our formalism and results, we also examine the potential presented in Ref.~\cite{Cyncynates:2023esj}:
\begin{align}
V(a,\,a_S)  =  m_a^2(T) \tilde{f}_a^2 \left[1-\cos\left(\frac{a_S}{\tilde{f}_S}+\frac{a}{\tilde{f}_a}\right) \right]  +\tilde m_S^2\tilde{f}_S^2\left[1-\cos\left(\frac{a_S}{\tilde{f}_S}\right)\right]\,.
\label{eq:VII}
\end{align}
As shown in Ref.~\cite{Murai:2024nsp}, with a unitary field redefinition this can be brought in the form
\begin{equation}
V(a,\,a_S)  =  m_a^2(T) F_a^2 \left[1-\cos\left(\frac{a}{F_a}\right)\right]  +M_S^2F_S^2\left[1-\cos\left(\frac{a_S}{F_S}+N_a\frac{a}{F_a}\right)\right]\,,
\end{equation}
where $F_a = \tilde{f}_a\tilde{f}_S/\sqrt{|\tilde{f}_S^2-\tilde{f}_a^2|},\, F_S = (\tilde{f}_S^2 +\tilde{f}_a^2)/\sqrt{|\tilde{f}_S^2-\tilde{f}_a^2|},\, M_S = \tilde m_S \tilde{f}_S \sqrt{|\tilde{f}_S^2-\tilde{f}_a^2|}/(\tilde{f}_a^2+\tilde{f}_S^2)$ and ${N_a = \tilde{f}_a^2/(\tilde{f}_S^2+\tilde{f}_a^2)}$. 
Expanding the potential to second order in the fields, we obtain the mass matrix
\begin{align}
\begin{pmatrix} 
a_S & a 
\end{pmatrix} 
\left(\begin{array}{cc}
M_S^2 & M_S^2\, N_a\,\frac{F_S}{F_a} \\
M_S^2\, N_a\, \frac{F_S}{F_a} & m_a^2(T)+ M_S^2\, N_a^2\, \frac{F_S^2}{F_a^2}
\end{array}\right) \begin{pmatrix} a_S \\  a \end{pmatrix} \,,
\end{align}
that can be matched to Eq.~\ref{eq:massMatrix} with
\begin{align}
    R_f = N_a \frac{F_S}{F_a} = \frac{\tilde{f}_a}{\tilde{f}_S}\,,\qquad R_m = \frac{M_S}{m_{a,0}} = \frac{\tilde m_s}{m_{a,0}}\tilde{f}_S\frac{\sqrt{|\tilde{f}_S^2-\tilde{f}_a^2|}}{\tilde{f}_a^2+\tilde{f}_S^2}\,.
\end{align} 
Therefore, the results obtained in Sec.~\ref{sec:PotentialI} apply and the LZ parameter reads
\begin{align}
\gamma =
\frac{M_S}{H(T_\times) n} \Big(N_a\frac{F_S}{F_a}\Big)^2 + \mathcal{O}\Big(\,N_a^4\frac{F_S^4}{F_a^4}\,\Big)\,.
\end{align}

\section{Dark matter abundances}~\label{sec:DMabundance}

With the knowledge of the previous sections, we now have a simple recipe to calculate the dark matter abundances in any given two-axion system. 
Regardless of whether the crossing is adiabatic or non-adiabatic, the evolution from the onset of axion oscillations up to just before the crossing is always adiabatic. Therefore, the relic density before the crossing can be computed by enforcing conservation of the adiabatic invariants, i.e., the independent comoving axion numbers, see \cref{eq:numberdensitiesWKB}. These are given by
\begin{gather}
   n_i^{\rm WKB} = \frac{\rho_i(T)R^3}{m_i(T)} \approx  \frac{1}{2} m_i(T_i^{\rm osc}) a_i (T_i^{\rm osc})^2  R ^3(T_i^{\rm osc}) . \label{eq:number densities, WKB}
\end{gather}
Taking Application I, section \ref{sec:PotentialI}, as an example, the initial field values of the mass eigenstates, $a_i(T_i^{\rm osc})$, can be found from Eq.~\eqref{Eq:InitialFieldValues}\,\footnote{\label{footnote on ToscH>Tx>ToscL}Note that Eq.~\eqref{Eq:InitialFieldValues} is only useful if there is no significant change in the mixing angle between $T_H^{\rm osc}$ and $T_L^{\rm osc}$. In this case, $a_L$ will remain static between $T_H^{\rm osc}$ and $T_L^{\rm osc}$ and we can calculate the initial field value from the assumed initial condition set at high temperatures. The scenario in which a significant change in the mixing angle takes place between $T_H^{\rm osc}$ and $T_L^{\rm osc}$ lies outside the range of validity of our analysis.}. To obtain the oscillation temperatures, $T_i^{\rm osc}$, we consider a refined condition 
\begin{gather}
    m_i(T_i^{\rm osc})\equiv x_i H(T_i^{\rm osc})\,,
    \label{eq:Tosc}
\end{gather}
where 
$x_i$ depends on the temperature-dependence of each mass eigenvalue. While 
the conventional choice $x\sim 3$ matches well the relic abundance produced by a QCD axion, a value of $x\sim$~1.6 produces a better fit to the abundance produced by a sterile field with a temperature-independent mass (see e.g. \cite{DiLuzio:2021gos}). In our two-axion scenario, this is complicated further by that fact that, depending on the input parameters, either of the states can develop a QCD-like or nearly-temperature-independent mass. To interpolate between the two cases, we perform a fit to Eq.~\ref{eq:Tosc} in a single-axion misalignment scenario with a generic mass $m_{i}\propto (T_{\rm QCD}/T)^{n'_i}$', where the parameter $n'_i$ depends on the mixing angle and interpolates between $0$ and $n$, as defined by Eq.~\eqref{eq:maofT}. We find that the numerical results are well-described by Eq.~\eqref{eq:Tosc} with the fit
\begin{gather}
    x_i = 1.6 + 0.6 n_i'\,.
\end{gather}
We use this condition to determine $T^{\rm osc}$ in our analytical solutions.

If the mixing angle varies slowly enough, $\dot \xi\ll \omega_{\rm osc}$, the evolution will be adiabatic during crossing and the comoving number densities of axions in~\eqref{eq:number densities, WKB} are conserved.
Thus, one simply evaluates these at $T=T_0$ to find the present day energy densities,
\begin{align}
    \label{eq:adiabaticEnergyDensity}
    \rho_i(T_0)=\frac{m_i(T_0) n_i^{\rm WKB}}{R_0^3}\,,
\end{align}
where the subscript `0' refers to present-day quantities. We emphasize that although this result looks standard, the non-standard evolution of $m_i(T)$ can lead to interesting effects.

As explained in \cref{sec:convprob}, when the mixing angle changes rapidly, $\dot \xi \gg \omega_{\text{osc}}$, the comoving axion number densities are no longer independently conserved, and a non-zero probability of conversion arises. The evolution in this non-adiabatic case can be described by the LZ probability for the transition between mass eigenstates.
As shown in Appendix \ref{app:Landau_Approach}, the LZ conversion probabilities modify the number densities after crossing according to
\begin{align}
n_H (T<T_{\times})  = {(1-P_{\rm LZ})} n_H^{\rm WKB} + {P_{\rm LZ}} n_L^{\rm WKB}\quad\text{and}\quad
n_L (T<T_{\times})  = {(1-P_{\rm LZ})}  n_L^{\rm WKB} + P_{\rm LZ}n_H^{\rm WKB}\,.
\label{eq:LZmod}
\end{align}
which reduces to Eq.\,\eqref{eq:adiabaticEnergyDensity} when $P_{\rm LZ} \rightarrow 0$, or equivalently, $\gamma \gg 1$.
The present day dark matter abundance carried by each eigenstate is then
\begin{align}
    \rho_i(T_0)=\frac{m_i(T_0) n_i(T<T_\times)}{R_0^3}\, .
\end{align}

Note that Eq.~\ref{eq:LZmod} implicitly assumes that the two fields start oscillating before the crossing temperature. Conversely, if $T_\times > T_{L,H}^{\rm osc}$, the changes in the mixing angle take place before the onset of oscillations and the scenario reduces to the standard WKB solution.
However, this does not necessarily result in a smooth matching prescription, and one should expect some error in the region where $T_{L,H}^{\rm osc}\sim T_\times $.

\subsection{Accuracy of the new analytical prescription}
\label{Sec:accuracy}

\begin{figure}[t]
\centering
\includegraphics[width=0.52\textwidth]{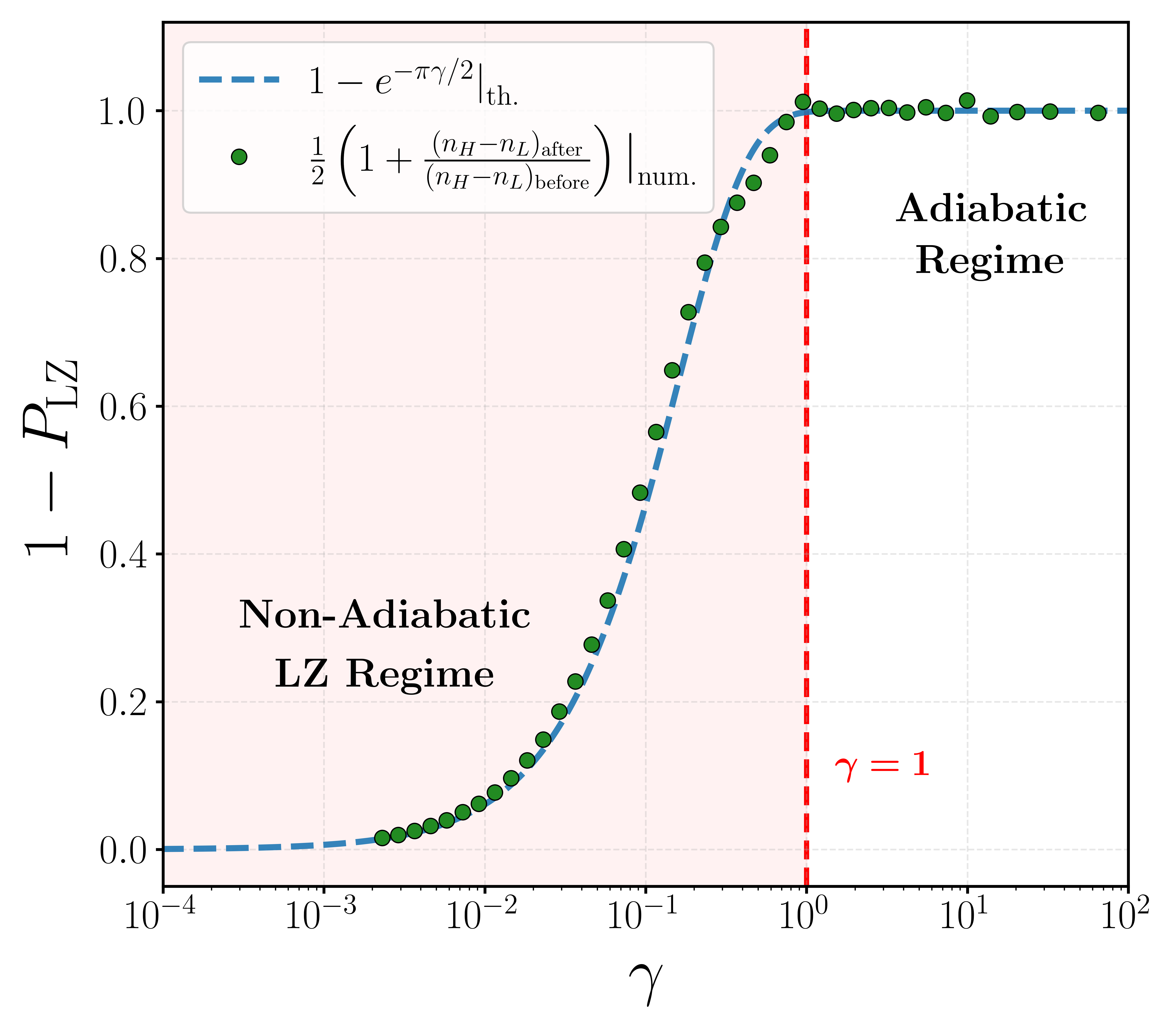}
\caption{\justifying Comparison of the analytical prediction of the LZ survival probability, $1 - P_{\rm LZ}$ (dashed blue line), and the results of the numerical simulation described in \cref{Sec:accuracy} (green points). We use the potential introduced in \cref{sec:PotentialI}, evaluated at $R_m = 4 \times 10^{-2}$ and $R_f \in [10^{-2},1]$.}
\label{fig:PLZ}
\end{figure}
To evaluate the accuracy of our analytic procedure, we have compared it against a full numerical solution using the two-field equations of motion. 
In Fig.~\ref{fig:PLZ}, we present the result of this comparison in the computation of the LZ survival probability, following~\cref{eq:LZProbability}. {We find that the analytical curve gives a very good fit to the numerical result, even in the very non-adiabatic regime. We here show a representative curve with $R_m = 4 \times 10^{-2}$ and $R_f \in [10^{-2},1]$.}

\begin{figure}[t]
\includegraphics[width=0.49\textwidth]{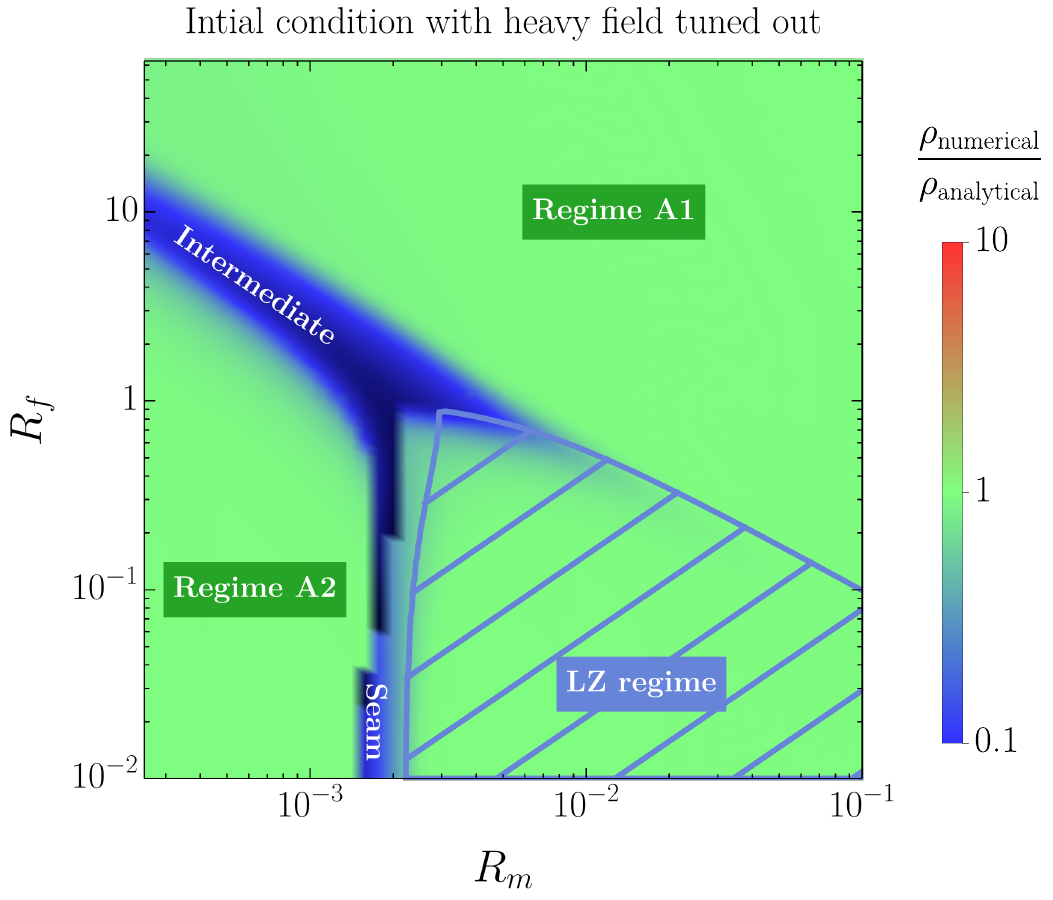}
\includegraphics[width=0.49\textwidth]{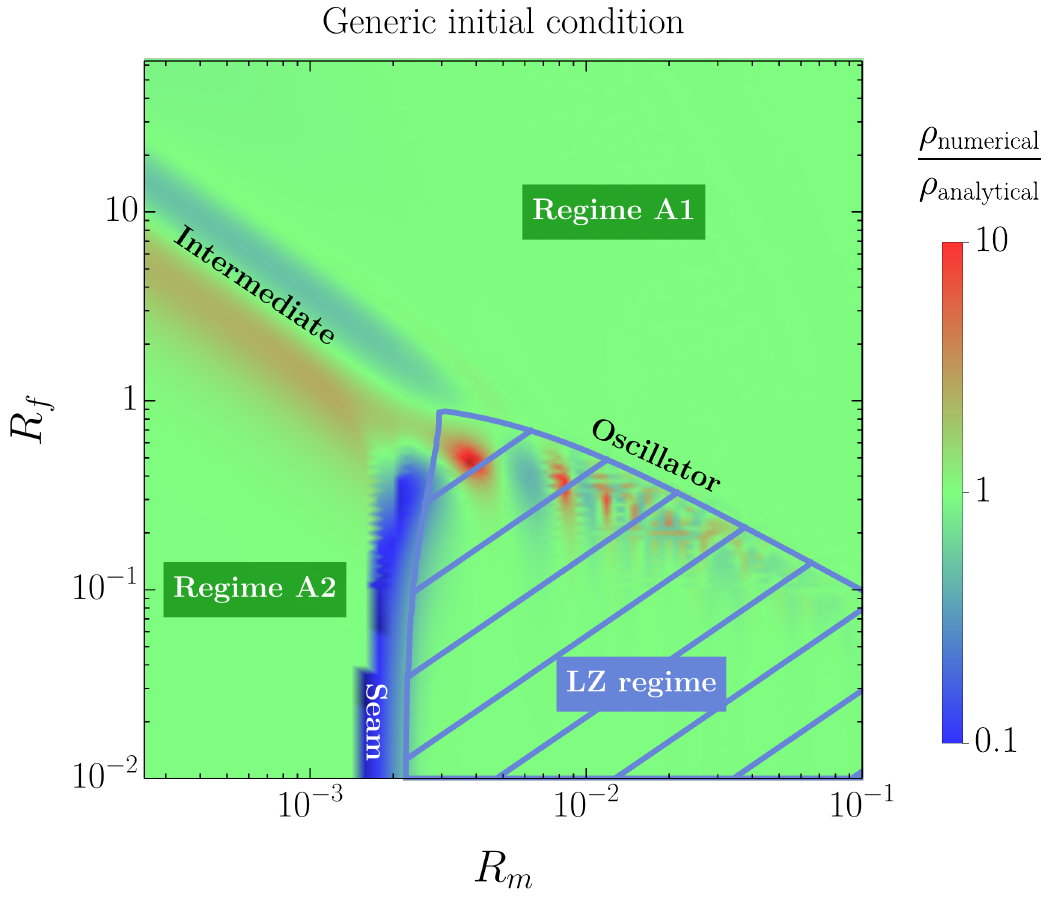}
\caption{\justifying
Evaluation of the precision of the analytic prescription. We show the ratio of total DM abundance predicted by the analytic and numerical prescriptions. \textbf{Left:} Discrepancy in a scenario in which the initial condition is tuned such that only the light field starts to oscillate, which corresponds to the results presented in \cite{Ho:2018qur}. 
\textbf{Right}: Discrepancy in a generic scenario without any special tuning. See section \ref{Sec:accuracy} for a discussion of the features.}
\label{fig:comparison}
\end{figure}

We now extend the discussion to the 2D parameter space $R_m$ vs $R_f$. Fig.~\ref{fig:comparison} shows the error in the computation of the total energy density using our analytical approach, relative to the numerical results.
It is evident that the two approaches generally agree, within an error of less than 10\%, both in the adiabatic and the LZ regions.
Nevertheless, some limitations of our prescription are visible, specially in the transitions between the LZ and the two adiabatic regions, labelled A1 and A2 in the plot, which are distinguished by whether the fields oscillate before or after $T_\times$ (see Fig. \ref{fig:masscross}).  We describe these in detail in the following.

In the region A2, the fields start oscillating only after $T_{\times}$ such that no significant change in the mixing angle takes place after the onset of oscillations.
Therefore, the fields evolve adiabatically in spite of $\gamma \ll 1$; see Fig.~\ref{fig:regionLC}. This behavior cannot be described continuously by the LZ formula, 
leading to the large errors in the region dubbed \textit{Seam}. 
Furthermore, even for $\gamma>1$, the fields can start to oscillate very close to $T_\times$. 
This leads to uncertainties in the definition of the initial misalignment angle upon the use of the WKB approximation, which results in the errors observed in the \textit{Intermediate} region. 
This scenario is depicted in the sketch of Fig.~\ref{fig:masscross}, along with other cases that illustrate the regimes represented in the regions A1 and A2.

Moreover, we observe a discrepancy along the $\gamma = 1$ line dividing the LZ and the A1 regions. 
In this regime, the heavy field behaves as a driven oscillator, whose amplitude is enhanced by the resonance in $\dot \xi$, around $t_\times$. Indeed, by solving the equation of motion~\eqref{eq:AxionAlpEOMMassBasisSimplified} neglecting $\ddot\xi$, we find that the heavy field is driven to an amplitude of order $a_H (t_\times) \approx (\sin{\gamma/\gamma}) a_L (t_\times) \cos{(m_H t_\times)}\to a_L (t_\times) \cos{(m_H t_\times)}$ in the instantaneous LZ approximation i.e. $\delta t_{\rm res} \ll 1/\omega_{\rm osc}$.
This demonstrates that~\textit{the LZ effect can be understood as the instantaneous limit of a driven oscillator}.

However, in the limit in which $\delta t_{\rm res}$ becomes comparable to the oscillation timescale, that is for $\gamma\to 1$,
the relative alignment of the phases of the two fields begins to have an effect.
This is beyond any analytical treatment that does not resolve the phase of the oscillations and therefore leads to the $\mathcal{O}(1)$ errors observed in the transition from LZ to the adiabatic region A1 shown on the right panel of Fig.~\ref{fig:comparison}.

While the dominant field is not sensitive to relative phases deep in the LZ regime, we remark that the subdominant field can pick up a phase-dependent error across the LZ parameter space (not visible in Fig.~\ref{fig:comparison}).
Therefore, while the total dark matter abundance is, apart from the discussed limitations, accurately predicted by our novel prescription, the abundance of the subdominant field should be taken as an $\mathcal O (1)$ estimate only. 

%
%
\newcommand{\ofT}{h}

\subsection{Extended parameter space for axion dark matter}
Based on our analytical formulas, we discuss in this section the extended parameter space that can open for axion dark matter in the presence of mass mixing, relative to that of a single QCD axion field with decay constant $f_a$, initial misalignment angle $\theta_a$ and energy density $\rho_a$. 
For simplicity, we focus on scenarios where a single field starts oscillating, i.e. $\theta_L = 0$ or $\theta_H =0$.

We consider, first, the potential in Sec.~\ref{sec:PotentialI} and the regime where $R_f,\,R_m \ll 1$. According to Eqs.~\eqref{eq:rot} and~\eqref{eq:angle_pot1}, in the high-temperature limit in which {the onset of oscillations} takes place, 
the light field behaves as the standard QCD axion ($\cos\xi \sim 0$), while the heavy field behaves as the sterile field. The opposite identification can be made at later times, i.e. $a_H\sim a$ while $a_L\sim a_S$.~A possible evolution of the eigenvalues representative of this case is depicted in  Fig.~\ref{fig:masscross}.
For $\theta_H=0$ and $\theta_L = \theta_a$, misalignment with {$m_L = m_a(T)$} then takes place and the light mass eigenstate is produced with a number density $n_L = n_a$. 
In the adiabatic regime, the number densities are conserved so that, at late times, this is the number density of a state with mass $\sim m_s$. Hence, the final energy density is expected to be modified by a factor $R_m$ relatively to that of a single QCD axion.
Contrarily, in the LZ regime (assuming that both fields oscillate before $T_\times$), the number densities of the light and heavy states are swapped (for $\gamma\ll 1$); therefore, at late times, the number density of the heavy eigenstate, with a mass $\sim m_{a} (T_0)$, becomes $n_H = n_a$. In this way, we expect no modifications with respect to the predictions of the single axion dark matter misalignment.

On the other hand, for $\theta_L = 0$ and $\theta_H = \theta_a$, only the heavy field oscillates and misalignment with $m = m_s$ occurs instead. The number density obtained in this case differs from the number density of a single QCD axion by a factor
\begin{gather}
    \frac{n_s}{n_a} = \frac{R_f^2}{\sqrt{R_m}} \sqrt{\ofT}\,, \qq{where} \ofT = m_a(T_{\rm osc})/m_a <1\,
    \label{eq:sterileMIS}
\end{gather}
and $n_s$ is the number density obtained for a sterile field with a mass and decay constant equal to $m_s$ and $f_s$, respectively.
In the adiabatic regime, the final energy density of the heavy field is then given by $n_s m_a$; while in the LZ regime, the final density (now associated to the light field) reads $n_s m_s$. 

These different limits are summarized in Tab.~\ref{tab:overview_lowRf_lowRM}. As it is apparent, the first tuned scenario with $\theta_L = \theta_a$ can 
either suppress or leave the dark matter abundance of a single QCD axion unchanged. Instead, the scenario with $\theta_H =\theta_a$ can enhance the single axion abundance if the fields evolve adiabatically with $R_m<R_f$. These simple estimates illustrate how much the single axion dark matter predictions can change in the presence of scalar mixing. Moreover, the very distinct results obtained in the adiabatic ($\gamma \sim 1$) and extremely non-adiabatic ($\gamma\ll 1$) scenarios illustrate the importance of employing $P_{\rm LZ}$ to accurately predict the energy densities in the parameter space that lies between these limits.
Other limits are trivial to compute, so we do not include them in the table. For instance, in the parameter space where $R_f\gg 1,\,R_m \ll 1$, the temperature dependence in Eq.~\ref{eq:massMatrix} and the mixing angle remain constant $(\cos\xi \sim 1)$. In this limit, there is no level crossing between the two axions (see Fig.~\ref{fig:regionLC}) and the individual number densities of the mass eigenstates are conserved.
This also applies to regimes where $R_m \gg 1$, even though the eigenstates have different decay constants in these two scenarios.
As noted before, similar conclusions apply to the model described in Sec.~\ref{sec:PotentialII} after matching the parameters in the two potentials.

\begin{table}[t]
\centering{}
	\renewcommand{\arraystretch}{2.9}
\setlength{\tabcolsep}{12pt} 
\begin{tabular}{c  c  c}
    \textbf{Scenario} & \textbf{Adiabatic} & \textbf{Non-adiabatic} \\ 
    \hline
   $\left(\theta_H = 0\,, \theta_L = \theta_a\right)$ & $\rho \sim R_m \rho_a$ & $ \rho \sim \rho_a$  \\
    \hline
     $\left(\theta_L = 0\,, \theta_H = \theta_a\right)$ 
     & $\rho \sim \sqrt{h} \frac{R_f^2}{\sqrt R_m} \rho_a$ & 
     $\rho \sim \sqrt{h} {R_f^2}{\sqrt R_m} \rho_a$ 
    \\
    \hline
\end{tabular}
\caption{\justifying
Analytical predictions for the dark matter relic abundance in tuned regimes. We have used as benchmark the potential in Sec.~\ref{sec:PotentialI} with $R_f,\, R_m \ll 1$. Here, $h = m_a(T_{\rm osc})/m_a <1$.
}
\label{tab:overview_lowRf_lowRM}
\end{table}

\section{Recast of haloscope bounds}\label{sec:haloscope}

An analytic understanding of the relic abundances of each axion field is crucial not only for identifying the masses and decay constants that can account for the entirety of dark matter, but also because the sensitivity of some of the most precise detection experiments (i.e. haloscopes) depends directly on these abundances. In particular, the signal power in a haloscope scales quadratically with the corresponding coupling and is proportional to the local axion dark matter density. Focusing on the axion-photon coupling,
\begin{align}
\mathcal{L}\supset \frac{1}{4}\big( g_{a_{H}\gamma\gamma} a_{H}+g_{a_{L}\gamma\gamma} a_{L}\big) F \widetilde{F} \,,
\end{align}
the signal power then scales as $\propto g_{a_i\gamma\gamma}^2\rho_{i}$.
The absence of a detected signal can thus be interpreted as an upper bound on  $g_{a\gamma\gamma}$, assuming that axions constitute 100\% of the dark matter. However, if an axion makes up only a fraction of the total dark matter content,  
 ${\rho_i}/{\rho_{\rm DM}}$,   
the experimental sensitivity to  $g_{a\gamma\gamma}$  is reduced by a factor  
$\sqrt{{\rho_i}/{\rho_{\rm DM}}}$. 

\begin{figure}[t]
\includegraphics[width=0.49\textwidth]{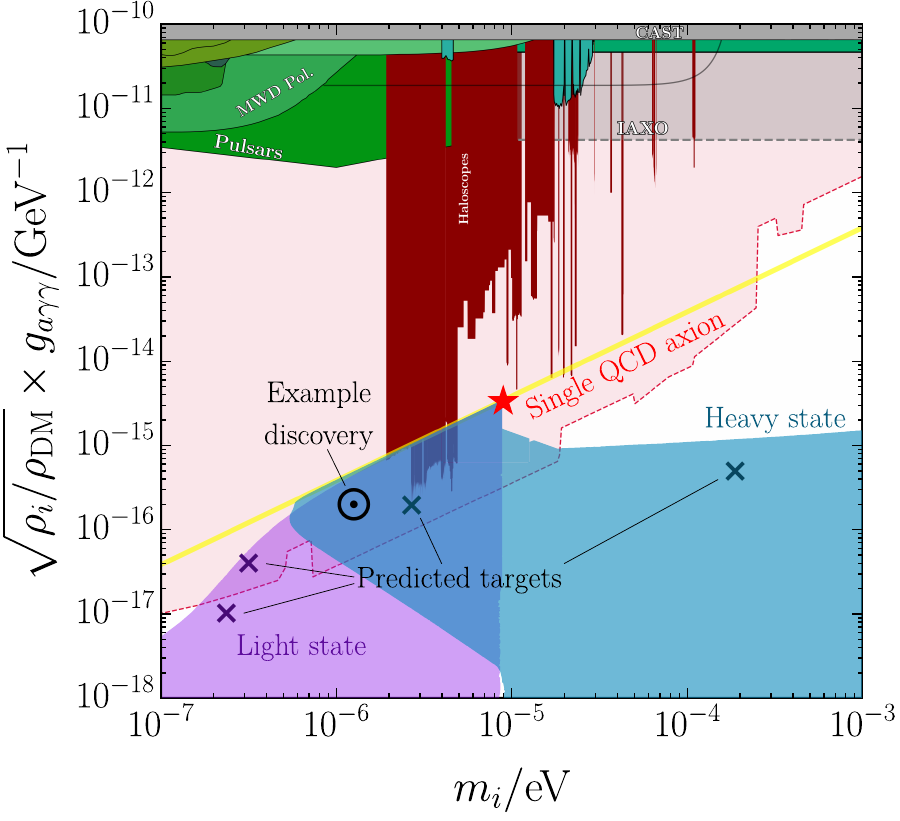}
\includegraphics[width=0.49\textwidth]{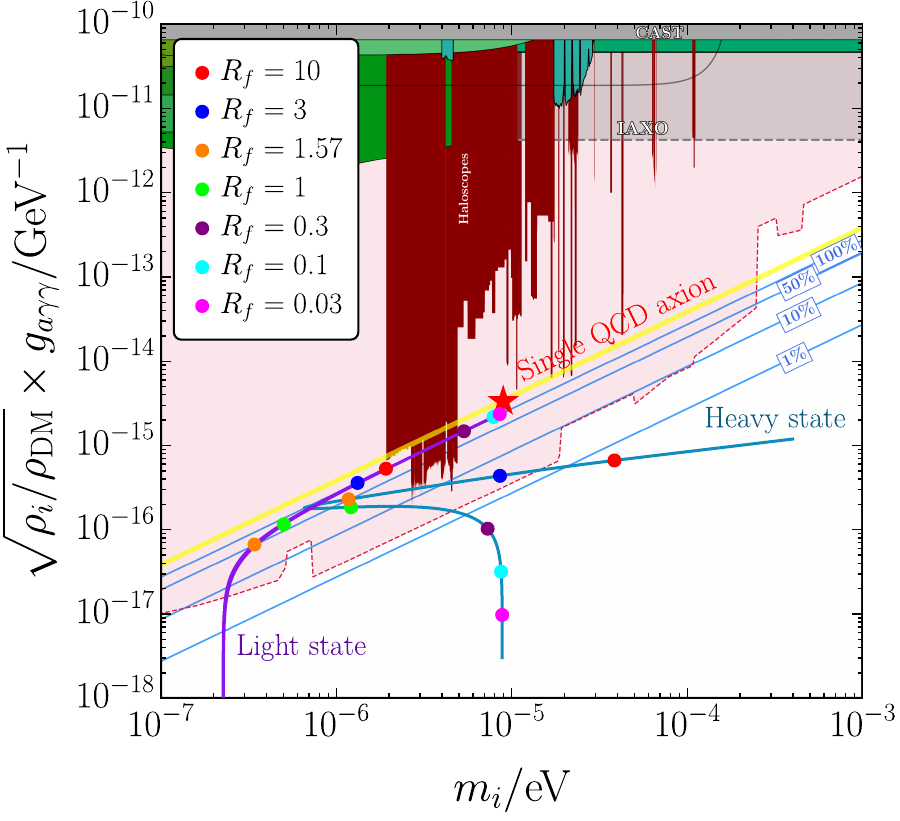}
\caption{\justifying 
\textbf{Left:} The region of parameter space that can be reached by two misaligned axions under assumption of the potential defined by \eqref{eq:VTAK} with $R_m\in \{10^{-3},\ 1$\} and $R_f \in \{10^{-5},\ 100\}$. We here assume a generic $\mathcal{O}(1)$ initial condition of $a/f_a=1.25$ and $a_s/f_s=0.75$ and no electromagnetic anomaly for each of the initial axion fields $E/N=0$. 
\textbf{Right:} The subset of parameter space in which Maxions can be found. Each colored pair of dots correspond to a solution with a given $R_f $. This gives an example of a concrete prediction of our results: If any one state is discovered in a detector, then our model implies the existence of the paired state, here indicated in dots of matching color.  Axion bounds adapted from Ref.~\cite{AxionLimits}{\hfill} 
}
\label{fig:Maxion 1.25}
\end{figure}

Taking this into account, we show in Fig.~\ref{fig:Maxion 1.25} (left panel) the regions of the parameter space where one could detect each of the two axions comprising together all of the dark matter in the Universe. For this plot, we consider the potential defined by Eq.~\eqref{eq:VTAK} assuming $\mathcal{O}(1)$ initial misalignment angles and no electromagnetic anomaly for each of the initial axion fields ($E/N=0$). Under this assumption, the coupling to photons arises only through mixing with the neutral mesons, 
{$g_{a_{L,H}\gamma\gamma}=-1.92\,\alpha_{\rm em}/(2\pi f_{L,H})$}, where the scales $f_{L,H}$ are defined so that the coupling to gluons of each mass eigenstate is $\frac{\alpha_s}{8\pi}\frac{{a_{L,H}}}{f_{L,H}}G\tilde G$.

Before commenting on the general expectations, let us discuss a specific model to localize explicitly each axion pair. 
Namely, in Fig.~\ref{fig:Maxion 1.25} (right panel), we 
show the position of the two axions that explain the totality of dark matter in
the \textit{QCD maxion} scenario presented in Ref.~\cite{Gavela:2023tzu}. In such scenario, the two eigenstates contribute equally to the solution to the strong CP problem and are therefore deviated from the canonical band by the same factor $\sqrt N = \sqrt 2$, i.e.
\begin{align}
m_H^2 f_H^2=m_L^2 f_L^2 = f_\pi^2 m_\pi^2 \frac{m_u m_d}{\left(m_u+m_d\right)^2} \times {2}\,,
\label{eq:maxionmass}
\end{align}
where the masses above refer to zero-temperature values.
Assuming the potential analysed in Sec.~\ref{sec:PotentialI}, these maximally deviated solutions are generated in the parameter space 
that satisfies~\cite{Gavela:2023tzu}
\begin{align}
\label{eq:maxion}
\text{tr}\,\mathbf{M}^2 = 2 m_a^2 \implies R_m^2 = \frac{1}{1+R_f^2}\,,
\end{align}
represented by a dashed blue line in Fig.~\ref{fig:regionLC}.
The maxion line does not intersect the LZ region unless $f_a > 10^{16}$\,GeV\,\footnote{We note that for other choices of the mixing potential, the maxion condition might not even intersect the LZ region at all.
For instance, in the scenario presented in Sec.~\ref{sec:PotentialII}, 
maxion eigenstates can never cross each other. This is easy to verify, as the maxion condition applied to the mass matrix in Eq.~\eqref{eq:VII}, 
$ R_m = \sqrt{1+ R_f^{-2}}$, 
occurs in the region where $m_a (T_{\times})>m_{a,0}$. }. For scales below this threshold, the evolution of the two maxion fields is therefore adiabatic.

The maxion condition in Eq.~\eqref{eq:maxion}
fixes one out of the three continuous parameters in the axion potential. In turn, the requirement that the two states make up for the totality of the dark matter observed restricts further the parameter space. We can therefore locate precisely the dark matter maxion pairs, as a function of $R_f$, as shown in Fig.~\ref{fig:Maxion 1.25}. 
Using Eq.~\ref{eq:maxionmass}, one can also simplify the expressions for the zero-temperature mass of the maxion fields and the decay constants $f_{L,R}$, setting the strength of the photon coupling:
\begin{align}
   & m_{L,H}^2 = m_{a,0}^2 \left(1\mp \frac{R_f}{\sqrt{1+R_f^2}}\right)\,;
   &  f_{L,H}^2 ={{2} f_{a}^2}{\left(1\mp \frac{R_f}{\sqrt{1+R_f^2}}\right)^{-1}} \,.
\end{align}
These show that the two fields become degenerate in mass in the $R_f\ll 1$ limit, sharing the same coupling to photons.

With these expressions in hand, we can carry an analysis similar to that of Tab.~\ref{tab:overview_lowRf_lowRM} to understand which field is expected to dominate the dark matter abundance in the various limits. In the regime $R_f\ll 1$, $\rho_L \sim R_f^{-2} \rho_H$; while in the opposite regime, $\rho_L \sim {R_f}^{3/2} \rho_H$. 
Therefore, in both scenarios, the light field is the dominant dark matter component, in agreement with the behaviour observed in the right panel of Fig.~\ref{fig:Maxion 1.25}. This is easy to understand: in both limits, the light field starts its oscillation with a much larger amplitude than the heavy one (see \cref{Eq:InitialFieldValues}), so it naturally carries most of the energy density.

Note also that in the first limit, $\rho_L\to \rho_a$ as well as $m_L\to m_a$. This explains the position of the upper pink dot, that converges to the position of the red star -- the benchmark for a single QCD axion which explains all the dark matter in the Universe ($\theta_i = \theta_a$) -- but with a slightly smaller coupling according to Eq.~\ref{eq:maxionmass}.
In the other limit, $R_f\gg 1$, $\rho_L \to {R_f}^{3/2} \rho_a$ and the light field misaligns with $m_a(T)$; this suggests that a larger $m_a$ is required to explain the dark matter abundance. However, since $m_L = m_a/R_f$, the dominant field is still found to the left of the red star.  

The most exotic signals we find in this scenario are obtained for $R_f\sim \mathcal O(1)$, where the two eigenstates share a comparable fraction of the dark matter energy density and can be deviated by one order of magnitude in mass relatively to the position of the red star. The crucial factor in this case is that for similar scales, the initial field values of the two axions are comparable; moreover, the larger mass of the heavy axion can compensate the enhancement of the energy density of the light field which misaligns with $m_L = m_a(T)$. 

\begin{figure}[t]
    \centering
    \includegraphics[width=0.49\textwidth]{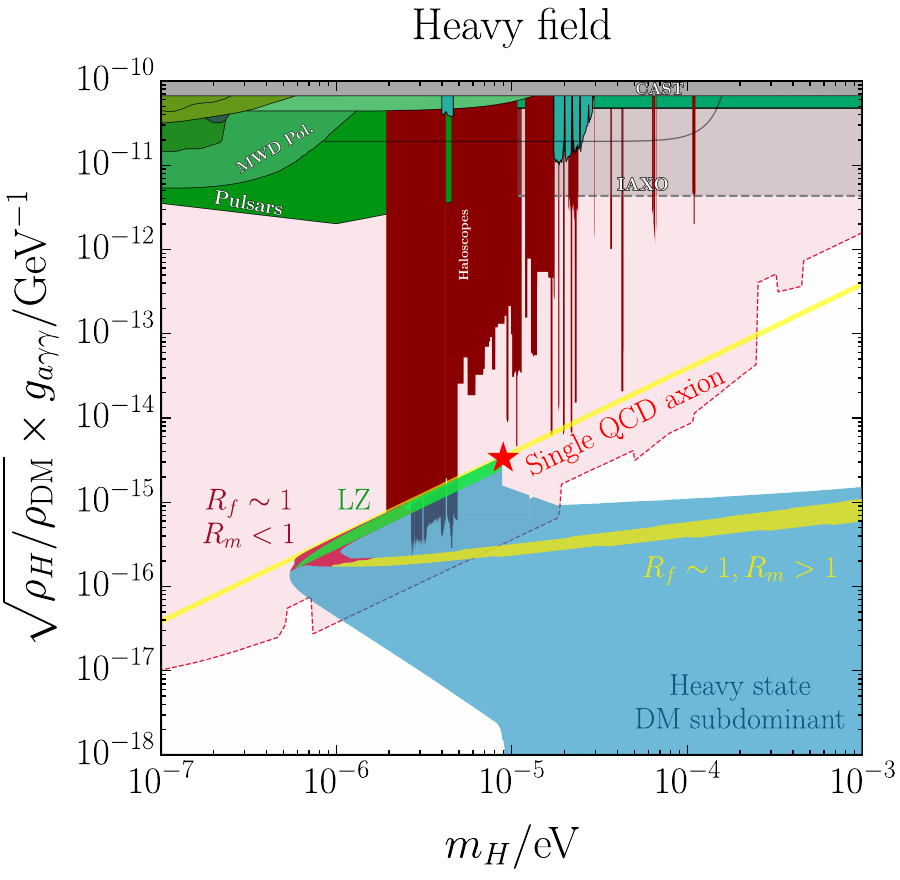}
    \includegraphics[width=0.49\textwidth]{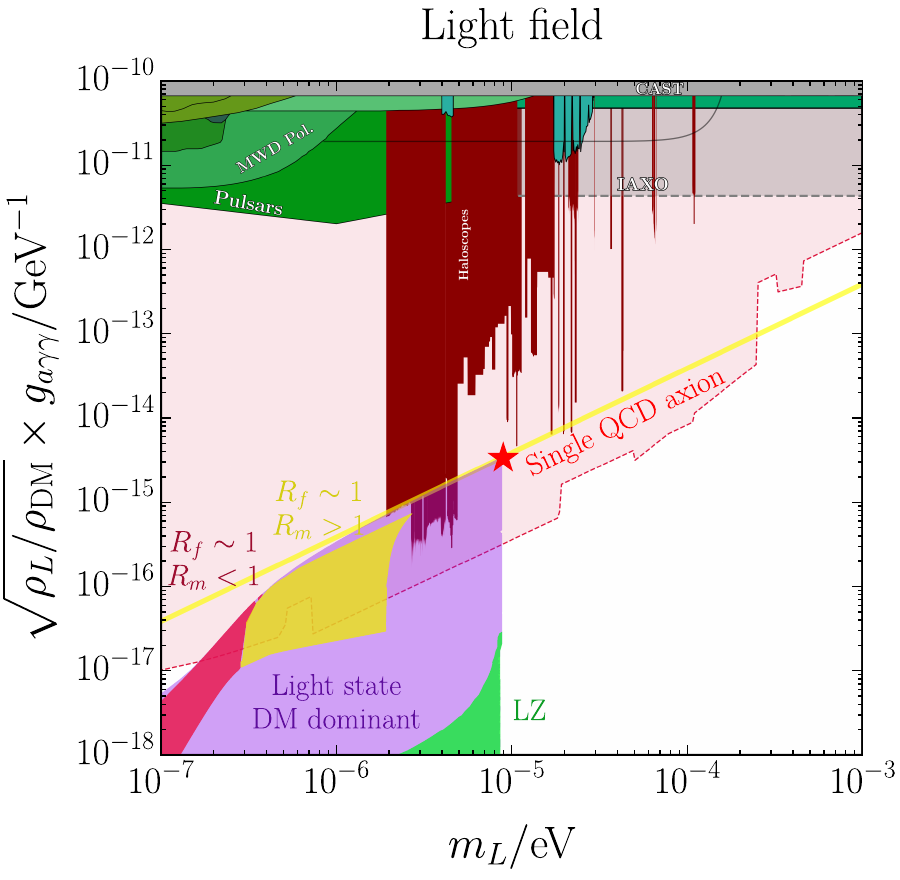}   
    \caption{\justifying
    \textbf{Left:} Regions where the heavy field, generically subdominant, explains the majority of dark matter in the Universe. Such regions open in three different scenarios: $R_f\sim \mathcal{O}(1)$ and $R_m < 1$ (red), $R_f\sim \mathcal{O}(1)$ and $R_m > 1$ (yellow) and the LZ regime (light green). \textbf{Right}: The same regions are highlighted in the parameter space occupied by the light field.}    
    \label{fig:dominationPlots}
\end{figure}

In more general scenarios that do not satisfy the maxion condition, as illustrated on the left plot in Fig.~\ref{fig:Maxion 1.25}, the 
initial amplitude of the heavy field is still suppressed relatively to that of the light field, unless $R_f \sim \mathcal O(1)$, see \cref{Eq:InitialFieldValues}. Therefore, if there is an adiabatic crossing, 
the light field will typically dominate the energy density
in a region far below the QCD axion line\,\footnote{Such region is absent in the QCD maxion scenario, where both fields are required to stay close to the canonical QCD line by forcing the condition in ~\cref{eq:maxion}.}, as illustrated in Fig.~\ref{fig:dominationPlots}. Interestingly, in these cases, the heavy field, identified with the QCD axion at zero temperature, can be found farther from the haloscopes reach as it represents only a small fraction of the dark matter energy density. Such setup has been previously identified in the literature~\cite{Ho:2018qur}.

The importance of our results lies in the quantitative understanding of the limits of the previous picture. Namely, if the parameters are such that the evolution of the two axions is non-adiabatic, the number density of the light field is converted into that of the heavy one. This opens a new region in the parameter space where crossing is possible, but the heavy dominates the energy density, for the same initial conditions. Such region is identified by green shading in Fig.~\ref{fig:dominationPlots}, close to the standard QCD axion line.

Although outside the range of the figure, in the regime $R_f > 1$ and $R_f\gg R_m^{-1}$, where mixing no longer modifies the cosmological evolution, the heavy field can also dominate the dark matter abundance as it misaligns as a sterile ALP with decay constant $f_a$ and mass $R_f m_S$, while $\rho_L \sim \rho_a$. 

Altogether, the new regions for QCD axion dark matter in Fig.~\ref{fig:Maxion 1.25} provide a guide to our understanding in case of a potential axion discovery to the right of the canonical band. For example, if a first signal would be found in the light purple region, one could have discovered the dominant dark matter component identified with the sterile field, while learning that the QCD axion should be found in the heavier mass region. 
Such a dominant field can have larger couplings relatively to what is obtained in the single ALP misalignment scenario, as the mixing with the QCD axion plays an important role in the cosmological evolution of the two fields, even if at zero-temperature they are decoupled.

In this $N=2$ axion scenario, one would be able to actually predict the location of a second signal if a first one was observed. This is illustrated with an example in the left panel of Fig.~\ref{fig:Maxion 1.25}, where we assumed that the first signal was found in the position marked with $\odot$. The dark matter requirement reduces the number of continuous parameters in the potential ($\{R_f,R_m,f_a\}$) to just two, so that the measurement of the coupling and mass of the first signal would fix these ratios. However, there is a two-fold ambiguity in identifying the second target, since in this specific example one would not know if the first signal corresponds to the heavy or the light axion; moreover, in some cases, a given value of the relic density can be obtained for different choices of $\{R_f,R_m,f_a\}$. For these reasons, in our example, a discovery of a signal at $\odot$ points to four distinct possible targets. Since they lie at different regions of sensitivity of haloscope experiments, the discovery of the complete axion system could require in this case a synergy of multiple experimental efforts.

Finally, we  remark that this predictive picture is only true under certain assumptions, most importantly that only the two axions comprise the totality of dark matter in the Universe and that they have the same model-dependent coupling to photons (in this case, different assumptions for $E/N$ simply represent a translation of the targets identified in Fig.~\ref{fig:Maxion 1.25}). Furthermore, the regions in this figure populated by the two axions can change for different mixing potentials. For example, as discussed in Sec.~\ref{sec:PotentialII}, the second scenario studied in this work (Eq.~\ref{eq:VII}) leads to the same LZ region as the previous one, but for different hierarchies of the axion decay constants, $R_f$. However, the initial field values of the two axions are different as well. For the potential~\ref{eq:VII}, non-adiabatic crossing occurs for $R_f\gg 1$ and $R_m\ll 1$, so that $a_H\gg a_L\sim a$ at high-temperature. Therefore, it is the light field that dominates the dark matter density in the LZ region, which means that the parameter space where dark matter can be explained far from the QCD axion line is now open due to the LZ effect.

\section{Conclusions}\label{sec:conclusions}

In this work, we have presented a novel analytical prescription to obtain the dark matter densities of a two-axion system in highly non-adiabatic regimes. This is based on the Landau-Zener approach to obtain the probabilities of conversion of the field number densities when the two eigenstates cross and the mixing angle changes abruptly such that its variation cannot be neglected in the evolution of the fields. In particular, we derive unambiguously the adiabaticity condition for the two axion system, which was only qualitatively discussed in the previous literature.

The implementation of the Landau-Zener approach to the axion misalignment is not straightforward: it relies on the linearization of the Klein-Gordon equation and, importantly, on the assumption that the diagonal elements in the flavor basis precisely match at the crossing temperature, i.e. that their difference evolves linearly in time during the entire timescale of the resonance. We have proven that this is true in the entire parameter space represented in Fig.~\ref{fig:regionLC} and beyond (up to particular tunings). An underlying assumption of our treatment is that the sum of number densities is conserved in the highly non-adiabatic scenarios, which we prove is an excellent approximation based on the linearized equations of motion of the two axions.

After a careful comparison between our novel analytical treatment and the numerical results, we found that -- apart from understandable limitations of our procedure, most striking when the fields start their oscillation very close to the crossing temperature -- we are able to predict the dark matter abundances of the two axion fields up to an error that is below 10\%. Our novel treatment can therefore be used to recast meaningfully the experimental bounds that rely on the assumption that a single signal is the totality of dark matter in the Universe.

Moreover, we show that under a potential discovery of a first signal to the right of the QCD axion band, one can now predict analytically a discrete set of targets where the second signal should lie such that both comprise all the dark matter observed. Such signals can lie in sensitivity windows of different haloscope experiments, such that a synergy of efforts might be necessary to discover the multiple axion systems.
We also find that a generic consequence of the mixing structures is that the number densities of the two eigenstates before the crossing are typically hierarchical (unless the two Peccei-Quinn VEVs are of similar magnitude). Particularly in these cases, the Landau-Zener conversion probability is fundamental to understand the regions populated by the dark matter field. Namely, the field carrying most of the energy density will be identified with the QCD axion or the sterile flavor depending on this probability. Even if the conversion is only partial, the growth in the population of the subdominant field can be such that it becomes visible to experiments.

The analytical results obtained in this work allow a clear comprehension of the novel regions that open for two axion dark matter models. This work also constitutes an important step towards the generalization of the misalignment mechanism to multiple axion dark matter scenarios. 
By treating an $N>2$ axion system as a series of two axion crossings, each of which can have a separate Landau-Zener conversion probability, our approach may be generalized to larger systems. The analysis of such systems will undoubtedly involve rich phenomenology, the exploration of which we leave to future work.

\section*{Acknowledgments}
C.A.M. was supported by the Office of High Energy Physics of the U.S. Department of Energy under contract DE-AC02-05CH11231. 
DD is supported by the James Arthur Postdoctoral Fellowship. 
The work of P.\,Q. is supported by the U.S. Department of Energy under grant number DE-SC0009919 and partially supported by the European Union's Horizon 2020 research and innovation programme under the Marie Sk\l odowska-Curie grant agreement No. 101086085-ASYMMETRY.
The work of P.S. was supported by the European Union – NextGeneration EU, mission 4, component 1, CUP C93C24004930006 and by the University of Padua via the MSCA Seal of Excellence @UNIPD 2024 programme, CUP C95F21010280001.

\newpage
\appendix

\section{Numerical analysis}
\newcommand{\ms}{m_S}
\newcommand{\as}{a_S}
\newcommand{\fs}{f_S}
\newcommand{\das}{\dot{a}_S}
\newcommand{\ddas}{\ddot{a}_S}
\newcommand{\thetas}{\theta_S}
\newcommand{\Rini}{R_{\rm ini}}
In this appendix, we detail how our numerical analysis was performed. This was used to cross check the analytic prescription provided in this paper. Specifically, we seek numerical solutions to the equations of motion generated by the the potential \eqref{eq:VTAK}, which are 
\begin{gather}
\begin{gathered}
	\ddot{a} + 3H\dot{a} + m_a^2(T) f_a \sin\left(\frac{a}{f_a}\right) + \ms^2\frac{\fs^2}{f_a} \sin\left(\frac{\as}{\fs} + \frac{a}{f_a}\right)  = 0, \label{eq:EOM, flavor states 1}
 \end{gathered}\\
	\ddas + 3H \das + \ms^2 \fs \sin\left(\frac{\as}{\fs}+\frac{a}{f_a}\right) = 0. \label{eq:EOM, flavor states 2}
\end{gather}
Since the rotation matrix, Eq. \eqref{eq:rot}, has a simple analytic solution in terms of temperature, it is possible to find the full solution for the mass eigenstates given numerical solutions for these flavor eigenstates.  
We can therefore choose to solve these equations at the level of the flavor states to avoid complications from the continuously changing mixing of the mass-eigenstates.

Solving such equations on cosmological timescales can lead to numerical problems because of the high number of oscillations. To improve robustness, we choose the variable
\begin{gather}
	u = \ln\left(\frac{R}{\Rini}\right)
\end{gather}
as our time variable~\cite{Karamitros:2021nxi}. Here $\Rini$ is the scale factor at the start of integration. Recasting the EOMs \eqref{eq:EOM, flavor states 1} and \eqref{eq:EOM, flavor states 2} in terms of this new time variable, the EOMs become
\begin{gather}
	\pddv{\theta}{u} + \left[3+\frac{1}{2}\pdv{\ln H^2}{u} \right]\pdv{\theta}{u}+\frac{m_a^2(T)}{H^2(T)}\sin{\theta} + \frac{\ms^2}{H^2(T)}R_f^2\sin(\theta+\thetas) = 0, \\
	\pddv{\thetas}{u} + \left[3+\frac{1}{2}\pdv{\ln H^2}{u} \right]\pdv{\thetas}{u}+ \frac{m_s^2}{H^2(T)}\sin(\theta+\thetas) =0,
\label{eq:EOM, recast}
\end{gather}
where $\theta=a / f_a$ and $\thetas = a_s / f_s $. The Hubble derivative can be conveniently expressed as 
\begin{gather}
	\pdv{\ln H^2}{u} = - \left(4+\pdv{\ln g_{*}}{\ln T}\right)\left(1+\frac{1}{3}\pdv{\ln g_{*s}}{\ln T}\right)^{-1},
\end{gather}
which is readily expressed as a tabulated function of scale factor \cite{Karamitros:2021nxi}. We use tabulated relativistic degrees of freedom provided by \cite{Saikawa:2018rcs}. However, we observe that taking full relativistic degrees of freedom into account reduces numerical stability. As our goal is to cross check our analytic solution, is sufficient to set $g_{*s}$ and $g_{*s}$ to a constant value while doing this comparison. After this verification, we then use the analytic solution with full relativistic degrees of freedom.

The recast EOM can then be solved directly by a numerical differential equation solving tool. We used \texttt{Mathematica} and \texttt{NDSolve}. As integration limits, we start the solver when either physical mass exceeds and $10^{-5} H$ and integrate until both physical masses exceed $3 \times 10^2 H $ and all change in mixing is complete. This yields full numerical solutions for $a(t)$ and $a_s(t)$, from which we analytically construct $a_H(t)$ and $a_L(t)$ as well as all derivative parameters such as $\rho(t)$ and $n(t)$.

Our numerical results demonstrate good agreement with our analytical prescription, verifying our solutions both in the adiabatic and in the LZ regimes. This results of the comparison are visualized in Fig. \ref{fig:comparison} and discussed in detail in Sec. \ref{Sec:accuracy}.

\section{Landau-Zener Formalism}\label{app:LZ}
In this section, we show the approximations and steps involved in going from the axion equation of motion \eqref{eq:AxionAlpEOM} to the Landau-Zener probability of conversion \eqref{eq:LZProbability}.
\subsection{Equation of Motion}
\label{app:FeshbachVillars}
After expanding the axion sine-gordon equation as the Klein-Gordon equation in \eqref{eq:AxionAlpEOM}, we define a comoving axion field $\tilde{a} = R(t)^n a$. The Klein-Gordon equation becomes
\begin{align}
    \label{eq:axionFullEOMAppendix}
    (\square + H(3- 2n) \partial_t + H^2n(n-1) + \mathbf{M}^2)
    \begin{pmatrix}
        \tilde{a}
        \\
        \tilde{a}_S
    \end{pmatrix}
    &= \mathbf{0}\,.
\end{align}
where $\square = \partial_t^2 - \nabla^2$ is the d'Alembertian operator and we assume a radiation-dominated universe such that $\ddot{R}/R = - H^2$. Choosing $n = 3/2$ eliminates the $\partial_t$ term, shifting the Hubble friction into a Hubble-dependent mass.
\begin{align}
    \label{eq:axionFullEOMAppendixEffectiveMass}
    (\square  + \mathbf{M_{\rm eff}}^2)
    \begin{pmatrix}
        \tilde{a}
        \\
        \tilde{a}_S
    \end{pmatrix}
    &= \mathbf{0}\,.
\end{align}
where $\mathbf{M}_{\rm eff}^2 = \mathbf{M}^2 + \frac{3}{4}H^2 \mathbb{1}$.
The axion equation of motion, \eqref{eq:axionFullEOMAppendixEffectiveMass} now takes the form of a two-state Klein-Gordon equation in flat space. This second-order differential equation can be equivalently re-written as two first-order differential equations by the technique of Feshbach and Villars \cite{feshbach1958elementary}. The advantage of this procedure is a direct mapping of the Klein-Gordon equation to the Schrödinger equation for non-relativistic axions, which can then be simply analyzed via the techniques of Landau and Zener.

The Feshbach-Villars decomposition proceeds as follows: let $\Psi = (\tilde{a} \quad \tilde{a}_S)^T$ be the comoving axion flavor doublet and $\Phi = \partial \Psi/\partial t$ be an auxiliary field.  With these definitions, \eqref{eq:axionFullEOMAppendixEffectiveMass} becomes
\begin{align}
    \label{eq:feshbachVillars}
    \partial_t \Phi = (\nabla^2  - \mathbf{M}_{\rm eff}^2)\Psi \, .
\end{align}
Next, we decompose $\Psi$ into a linear combination $\Psi = \Psi_+ + \Psi_-$ with 
\begin{align}
    \label{eq:feshbachVillarsChargeConjugate}
    \Psi_{\pm} \equiv \frac{1}{2}(\Psi  \pm i \mathbf{M}^{-1}_{\rm eff} \Phi) \, .
\end{align}
We will shortly see that the $\pm$ subscript indicates  positive and negative frequency modes of the axion field associated with particles and antiparticles; since the axion field is real, or equivalently, the axion is its own antiparticle, both modes must be accounted for.

Taking the time derivative of Eq.\,\eqref{eq:feshbachVillarsChargeConjugate} and inserting Eq.\,\eqref{eq:feshbachVillars} for $\partial_t \Phi$ gives a coupled pair of first-order differential equations (ie Schrödinger equations) \textit{mathematically equivalent} to Eq.\,\eqref{eq:axionFullEOMAppendixEffectiveMass}
\begin{align}
    \label{eq:feshbachVillarsSchrodingerFlavor}
    i\partial_t \Psi_{\pm} =& \pm[\mathbf{M}_{\rm eff} \Psi_{\pm} - \frac{\mathbf{M}_{\rm eff}^{-1}}{2}\nabla^2(\Psi_+ + \Psi_-) -\frac{i}{2}\mathbf{M}_{\rm eff}^{-1} (\partial_t\mathbf{M}_{\rm eff}) (\Psi_+ - \Psi_-)].
\end{align}
Note that although the $\frac{1}{2}\mathbf{M}_{\rm eff}^{-1}\nabla^2$ makes Eq.\,\eqref{eq:feshbachVillarsSchrodingerFlavor} look like a \textit{non-relativistic} Schrödinger equation, it is in fact completely relativistic.

In deriving Eq.\,\eqref{eq:feshbachVillarsSchrodingerFlavor}, we have made use of the matrix calculus identity $\partial_t \mathbf{M}^{-1} = - \mathbf{M}^{-1} (\partial_t \mathbf{M}) \mathbf{M}^{-1}$ for any matrix $\mathbf{M}$.
While Eq.\,\eqref{eq:feshbachVillarsSchrodingerFlavor} is in the axion flavor basis, it is convenient to transform to the instantaneous mass basis. Because there exists a unitary operator $U(\xi(t))$ that instantaneously diagonalizes  $\mathbf{M}_{\rm eff}^2$ at every time $t$ such that $\mathbf{M}_D = U^\dagger \mathbf{M}_{\rm eff} U$ is diagonal, the fields 
\begin{align}
    \label{eq:massEigenstates}
    {\Psi}_{D+} = U^\dagger {\Psi}_+ = 
    \begin{pmatrix}
        \tilde{a}_{H_+}
        \\
        \tilde{a}_{L_+}
    \end{pmatrix}\,,
    \qquad
    {\Psi}_{D-} = U^\dagger {\Psi}_- = 
    \begin{pmatrix}
        \tilde{a}_{H_-}
        \\
        \tilde{a}_{L_-}
    \end{pmatrix}
\end{align}
are the comoving axion mass eigenstates with positive and negative frequencies, respectively. In terms of this mass basis, Eq.\,\eqref{eq:feshbachVillarsSchrodingerFlavor} can be decomposed into a sum of  diagonal and non-diagonal terms (with respect to the two $\tilde{a}_H, \tilde{a}_L$ fields) such that 
\begin{align}
    \label{eq:feshbachVillarsSchrodingerMass}
    i\partial_t \Psi_{D,\pm} =& (H_D + H_{ND}) \Psi_{D,\pm}  \, ,
\end{align}
where the diagonal Hamiltionian is
\begin{align}
    H_D \Psi_{D,\pm} =& \pm [\mathbf{M}_{D} \Psi_{D,\pm} - \frac{\mathbf{M}_{D}^{-1}}{2}\nabla^2(\Psi_{D,+} + \Psi_{D,-})  -\frac{i}{2}\mathbf{M}_{D}^{-1} \partial_t\mathbf{M}_{D} (\Psi_{D+} - \Psi_{D-})] \, ,
\end{align}
and the non-diagonal Hamiltonian is
\begin{align}
    H_{ND}\Psi_{D,\pm} =& -i\Big[U^\dagger \dot{U}\Psi_{D,\pm} \pm \frac{1}{2} \big(\mathbf{M}_{D}^{-1}
   U^\dagger \dot{U} \mathbf{M}_{D} + \dot{U}^\dagger U\big)(\Psi_{D+} - \Psi_{D-})\Big].
\end{align}

Taking $\Psi_D$ to be momentum eigenstates such that $-i\nabla \Psi_D(\mathbf{p}) = \mathbf{p} \Psi_D(\mathbf{p})$
\footnote{Since the focus of this paper is on pre-inflationary axions systems such that cosmological production of axion dark matter is produced solely via misalignment, we will eventually take the non-relativistic limit with $\mathbf{p} \rightarrow \mathbf{0}$.}, 
the full Hamiltonian, Eq.\,\eqref{eq:feshbachVillarsSchrodingerMass} can be explicitly written out as
\renewcommand{\arraystretch}{2.0}
\newcommand\SmallMatrix[1]{{%
  \arraycolsep=0.35\arraycolsep\ensuremath{\begin{pmatrix}#1\end{pmatrix}}}}
\begin{align}
    \label{eq:FVHamiltonian}
    i\partial_t
     \SmallMatrix{
        \tilde{a}_{H_+}
        \\
        \tilde{a}_{L_+}
        \\
        \tilde{a}_{H_-}
        \\
        \tilde{a}_{L_-}
    }
    &=
    \SmallMatrix{
        m_H + \frac{\mathbf{p}^2}{2m_H} - \frac{i}{2} \frac{\dot{m_H}}{m_H} 
        && 
        i \dot{\xi} - \frac{i(m_H-m_L)\dot{\xi}}{2m_H}
        && 
        \frac{\mathbf{p}^2}{2m_H} + \frac{i}{2} \frac{\dot{m_H}}{m_H}
        &&
        \frac{i(m_H-m_L)\dot{\xi}}{2m_H}
        \\
        -i \dot{\xi} - \frac{i(m_H-m_L)\dot{\xi}}{2m_L}
        &&
         m_L + \frac{\mathbf{p}^2}{2m_L} - \frac{i}{2} \frac{\dot{m_L}}{m_L} 
        &&
        \frac{i(m_H-m_L)\dot{\xi}}{2m_L}
        &&
        \frac{\mathbf{p}^2}{2m_L} + \frac{i}{2} \frac{\dot{m_L}}{m_L}
        \\
        -\frac{\mathbf{p}^2}{2m_H} + \frac{i}{2} \frac{\dot{m_H}}{m_H} 
        &&
        \frac{i(m_H-m_L)\dot{\xi}}{2m_H}
        &&
        -m_H - \frac{\mathbf{p}^2}{2m_H} - \frac{i}{2} \frac{\dot{m_H}}{m_H} 
        &&
        i \dot{\xi} - \frac{i(m_H-m_L)\dot{\xi}}{2m_H}
        \\
        \frac{i(m_H-m_L)\dot{\xi}}{2m_L}
        &&
        -\frac{\mathbf{p}^2}{2m_L} + \frac{i}{2} \frac{\dot{m_L}}{m_L}
        &&
        -i \dot{\xi} - \frac{i(m_H-m_L)\dot{\xi}}{2m_L}
        &&
        -m_L - \frac{\mathbf{p}^2}{2m_L} - \frac{i}{2} \frac{\dot{m_L}}{m_L} 
   }
    \SmallMatrix{
        \tilde{a}_{H_+}
        \\
        \tilde{a}_{L_+}
        \\
        \tilde{a}_{H_-}
        \\
        \tilde{a}_{L_-}
    }
    \, .
\end{align}
In the extreme adiabatic limit where the terms $\dot{m}_H/m_H, \dot{m}_L/m_L$, and $\dot{\xi}$ are much smaller than $m_H$ and $m_L$ the four eigenvalues of Eq.\,\eqref{eq:FVHamiltonian} yield the standard relativistic dispersion relations,
\begin{align}
    E = \{\pm \sqrt{\mathbf{p}^2 + m_H^2}, \pm \sqrt{\mathbf{p}^2 + m_L^2}\}  \qquad (\text{Extreme adiabatic eigenvalues})\, ,
\end{align}
with $\pm$ corresponding to the positive/negative frequency modes, respectively.

In general, however, the terms $\dot{m}_H/m_H, \dot{m}_L/m_L$, and $\dot{\xi}$ cannot be neglected, especially in the extreme non-adiabatic regime where $\dot{\xi}$ can be much greater than $m_H$ and $m_L$. When keeping only the leading order non-adiabatic  ($\dot{m}_H/m_H, \dot{m}_L/m_L$) and kinetic ($\mathbf{p}^2$) corrections to adiabatic propagation and  mass-eigenstate mixing ($i\dot{\xi})$
\footnote{The off-diagonal term $i\dot{\xi}(m_H-m_L)/m_{H,L}$ can be neglected since it is always smaller than $i \dot{\xi}$ when mass-eigenstate mixing becomes important near the resonance point when $\sin^2 2\xi(t_\times) = 1$ and $\dot{\xi} \gg m_{H} \simeq m_L$. This follows from Eq.\,\eqref{eq:tan2xi} which indicates that on resonance, $\dot{\xi}(m_H-m_L) \lesssim m_{H,L}^2$.}
, the full $4\times4$ Hamiltonian becomes block-diagonal, splitting into the following two $2\times2$ Hamiltonians,

\renewcommand{\arraystretch}{2.0}
\begin{alignat}{3}
    \label{eq:FVHamiltonianApproxPlus}
    i\partial_t
     \SmallMatrix{
        \tilde{a}_{H_+}
        \\
        \tilde{a}_{L_+}
    }
    &\simeq
    \SmallMatrix{
        m_H + \frac{\mathbf{p}^2}{2m_H} - \frac{i}{2} \frac{\dot{m_H}}{m_H} 
        && 
        i \dot{\xi} 
        \\
        -i \dot{\xi}
        &&
         m_L + \frac{\mathbf{p}^2}{2m_L} - \frac{i}{2} \frac{\dot{m_L}}{m_L} 
   }
    \SmallMatrix{
        \tilde{a}_{H_+}
        \\
        \tilde{a}_{L_+}
    }
    \quad&\equiv & \quad \mathcal{H}_+ \Psi_{D+}
    \\
    \label{eq:FVHamiltonianApproxMinus}
    i\partial_t
     \SmallMatrix{
        \tilde{a}_{H_-}
        \\
        \tilde{a}_{L_-}
    }
    &\simeq
    \SmallMatrix{
        -m_H - \frac{\mathbf{p}^2}{2m_H} - \frac{i}{2} \frac{\dot{m_H}}{m_H} 
        && 
        i \dot{\xi}
        \\
        -i \dot{\xi}
        &&
         -m_L - \frac{\mathbf{p}^2}{2m_L} - \frac{i}{2} \frac{\dot{m_L}}{m_L} 
   }
    \SmallMatrix{
        \tilde{a}_{H_-}
        \\
        \tilde{a}_{L_-}
    }
   \quad&\equiv & \quad \mathcal{H}_- \Psi_{D-}
    \, .
\end{alignat}
In the non-relativistic and adiabatic limit, the solutions of Eq.\,\eqref{eq:FVHamiltonianApproxPlus} and \eqref{eq:FVHamiltonianApproxMinus} become
\begin{alignat}{3}
    \label{eq:aHAdiabaticSolution}
    \tilde{a}_{H \pm} &= \exp\left[\mp i \int_{t_{\rm osc}}^t m_H(\tau) d\tau  - \frac{1}{2} \int_{t_{\rm osc}}^t \frac{d \ln m_H}{d\tau}d\tau \right] \quad &=\quad &\tilde{a}_{H\pm}(t_{\rm osc}) \sqrt{\frac{m_H(t_{\rm osc})} {m_H(t)}} \exp[\mp i \int_{t_{\rm osc}}^t m_H(\tau) d\tau] \, ,
    \\
    \label{eq:aLAdiabaticSolution}
    \tilde{a}_{L \pm} &= \exp\left[\mp i \int_{t_{\rm osc}}^t m_L(\tau) d\tau  - \frac{1}{2} \int_{t_{\rm osc}}^t \frac{d \ln m_L}{d\tau}d\tau \right] &=\quad &\tilde{a}_{L\pm}(t_{\rm osc}) \sqrt{\frac{m_L(t_{\rm osc})}{m_L(t)}} \exp[\mp i \int_{t_{\rm osc}}^t m_L(\tau) d\tau] \, .
\end{alignat}
Since $\Psi = \Psi_{+} + \Psi_-$, the full solution is
\begin{alignat}{3}
    &\tilde{a}_H(t) = \tilde{a}_{H+}(t) + \tilde{a}_{H-}(t) \quad &=\quad & \tilde{a}_H(t_{\rm osc})\sqrt{\frac{{m_{H,\text{osc}}}}{{m_H}}}  \cos\left(\int^t_{t_\text{osc}} m_H(\tau) \mathrm{d}\tau\right) \, ,
    \\
    &\tilde{a}_L(t) = \tilde{a}_{L+}(t) + \tilde{a}_{L-}(t) \quad &=\quad & \tilde{a}_L(t_{\rm osc})\sqrt{\frac{{m_{L,\text{osc}}}}{{m_L}}}  \cos\left(\int^t_{t_\text{osc}} m_L(\tau) \mathrm{d}\tau\right) \, .
\end{alignat}
which, after recalling $\tilde{a}(t) \equiv a(t)R(t)^{3/2}$ is the comoving axion field, reduces to the standard WKB solution of the Klein-Gordan equation of Eq.\,\eqref{Eq:WKBexactsolmain}. Above, the coefficients $\tilde{a}_{H+}(t_{\rm osc}) = \tilde{a}_{H-}(t_{\rm osc})^*= \tilde{a}_H(t_{\rm osc})/2$.

As demonstrated by Eqns.\,\eqref{eq:aHAdiabaticSolution} and \eqref{eq:aLAdiabaticSolution}, the four separate quantities $|m_H\tilde{a}_{H\pm}^2|$ and $|m_L\tilde{a}_{L\pm}^2|$ are conserved in  adiabatic evolution, corresponding to the comoving number densities of each axion mode. Even in non-adiabatic evolution, however, there remain conserved quantities: Noting that $ \partial_t(\Psi_{D\pm}^\dagger\Psi_{D\pm}) = -i \Psi_{D\pm}^\dagger(\mathcal{H}_{\pm} - \mathcal{H}_{\pm}^\dagger)\Psi_{D\pm} $ where $\mathcal{H}_{\pm}$ is the $2\times2$ Hamiltonian of Eqns.\,\eqref{eq:FVHamiltonianApproxPlus} and \eqref{eq:FVHamiltonianApproxMinus}, implies that for all time,
\begin{align}
    \label{eq:probabilityConservation}
    \partial_t|\tilde{a}_{H\pm}|^2 + \partial_t|\tilde{a}_{L\pm}|^2 + |\tilde{a}_{H\pm}|^2\partial_t \ln m_H + |\tilde{a}_{L\pm}|^2\partial_t \ln m_L
 = 0\,.\end{align}
For any brief period of time, such as near a non-adiabatic resonance, the last two terms of Eq.\,\eqref{eq:probabilityConservation} can be neglected since the timescale of the resonance, $\delta t_{\rm res}$ is much shorter than the timescale for which the masses vary, which is $\mathcal{O}(H(t_\times))$, and since the field values themselves, $\tilde{a}_{H,L}$, change rapidly. In this regime, Eq.\,\eqref{eq:probabilityConservation} can be interpreted as a conservation of probability: $ \partial_t|\tilde{a}_{H\pm}|^2 + \partial_t|\tilde{a}_{L\pm}|^2 \simeq 0$. From here on, when quantitatively analyzing the behavior of the axion fields around the resonance time $t_{\times}$, we will drop the logarithmic correction to the masses $\dot{m}_{H,L}/m_{H,L}$. 

In the following two sections, we show two different ways of determining the evolution of the axion fields during the time $- \delta t_{\rm res} + t_{\times} \lesssim t \lesssim t_{\times} + \delta t_{\rm res}$ (see Fig.\,\ref{fig:masscross}), corresponding to Landau's approach (\ref{app:Landau_Approach}), and Zener's approach (\ref{app:Zener_Approach}).  These two methods determine the value of the axion fields before and after the resonance, thereby connecting the two WKB evolutions valid on opposite sides of the resonance. As a whole, this procedure provides an analytic determination of the asymptotic axion yields which may make up the dark matter.

\subsection{Landau's Approach}
\label{app:Landau_Approach}
Landau's method for determining the probability of conversion between two mass eigenstates after a non-adiabatic resonance involves  analytically continuing the adiabatic equation of motion along a path in the complex plane that bypasses the resonance time while always remaining adiabatic.

To see how this is done, first expand the effective $2\times2$ Hamiltonian $\mathcal{H}$ (Eqns.\,\eqref{eq:FVHamiltonianApproxPlus} and \eqref{eq:FVHamiltonianApproxMinus}) in terms of symmetric and antisymmetric diagonal pieces (equivalently, a linear combination of $\mathbb{1}$, $\mathbf{\sigma}_z$)
\begin{align}
    \label{eq:schrodingerEqFeshbachVillars2}
    i \partial_t
    \begin{pmatrix}
        \tilde{a}_{H \pm}
        \\
        \tilde{a}_{L \pm}
    \end{pmatrix}
    &= 
    \pm
    \begin{pmatrix}
        \frac{m_H + m_L}{2} && 0
        \\
        0 &&  \frac{m_H + m_L}{2}
    \end{pmatrix}
    \begin{pmatrix}
        \tilde{a}_{H \pm}
        \\
        \tilde{a}_{L \pm}
    \end{pmatrix}
    +
    \begin{pmatrix}
        \pm\frac{m_H - m_L}{2} - \frac{i}{2}\frac{\dot{m_H}}{m_H}  &&  i \dot{\xi}
        \\
        -i \dot{\xi} &&  \mp\frac{m_H - m_L}{2}  - \frac{i}{2}\frac{\dot{m_L}}{m_L} 
    \end{pmatrix}
    \begin{pmatrix}
        \tilde{a}_{H \pm}
        \\
        \tilde{a}_{L \pm}
    \end{pmatrix} \, .
\end{align} 
The first, or symmetric term in Eq.\,\eqref{eq:schrodingerEqFeshbachVillars2} is just a phase mutual to both $\tilde a_{H,L}$, and hence can be removed by temporarily redefining the axion fields by $\tilde{a}_{\pm} \rightarrow \tilde{a}_{\pm} \exp[\pm i \int_{t_0}^t (m_H + m_L)/2]$. The second, or antisymmetric term is physical. 
For notational brevity, we hereafter write just the solutions to the positive modes, $\tilde{a}_{H+}$ and $\tilde{a}_{L+}$; the negative modes, $\tilde{a}_{H-}$ and $\tilde{a}_{L-}$ proceed in the same manner. 

According to the equation of motion, Eq.\,\eqref{eq:schrodingerEqFeshbachVillars2}, when $\dot{\xi} \ll m_H -m_L$, 
the axion fields evolve adiabatically as 
\begin{align}
    \label{eq:adiabaticSolution}
    \tilde{a}_{H+}(t) &= \tilde{a}_{H+}(t_0)\sqrt{\frac{m_H(t_0)}{m_H(t)}}\exp\left[-i \int_{t_0}^t  dt' \frac{m_H(t')-m_L(t')}{2}\right]
    \\
    \tilde{a}_{L+}(t) &= \tilde{a}_{L+}(t_0)\sqrt{\frac{m_L(t_0)}{m_L(t)}}\exp\left[+i \int_{t_0}^t  dt' \frac{m_H(t')-m_L(t')}{2}\right]\,.
\end{align}
Eq.\,\eqref{eq:adiabaticSolution} is a good approximation when a resonance or level-crossing has not occurred, or equivalently, when $t$ in the integral of \eqref{eq:adiabaticSolution} is $\ll$ than the level-crossing time, $t_\times$ (see Eq.\, \eqref{eq:tCross}). 

Now, consider what happens when a resonance between the two axion fields occurs at time $t_\times$. Eq.\,\eqref{eq:adiabaticSolution}  assumes the axion evolution is adiabatic, which \textit{is not true} at $t_\times$. Nevertheless, Eq.\,\eqref{eq:adiabaticSolution} can \textit{still} be used to solve the behavior of the axion system around $t_{\times}$ by integrating $t$ along a contour in the complex plane that is always adiabatic, a trick due to Landau \cite{Landau:1932vnv,landau1977quantum} (see also \cite{Pizzochero:1987fj,kuo1989nonadiabatic} for elaboration on Landau). The contour is shown in Fig.\,\ref{fig:LZComplexPlane}.

This can be seen by first noting that the integrand of Eq.\,\eqref{eq:adiabaticSolution} can be written as $m_H-m_L = (m_H^2-m_L^2)/(m_H+m_L)$ where
\begin{align}
    \label{eq:massSqDifference}
    m_H^2-m_L^2 = \sqrt{(m_{aa}^2 - m_{ss}^2)^2 + 4 (m_{as}^2)^2} \, ,
\end{align}
which follows from Eq.\,\eqref{eq:tan2xi}.
Eq.\,\eqref{eq:massSqDifference} demonstrates that $m_H^2 - m_L^2$ (and hence $m_H - m_L$) is never zero for any \textit{real} time. At the resonant time $t_\times$, $m_{aa}^2 - m_{ss}^2 = 0$,  and the two mass eigenvalues are at their closest approach in real time, though never actually degenerate (so-called \textit{avoided level-crossing}). In the complex plane, however, there \textit{is} an actual level-crossing. To see this, Taylor expand $m_{aa}^2$ and $m_{ss}^2$ about $t_{\times}$:
\begin{align}
    m_{aa}^2 - m_{ss}^2 &\simeq  m_{aa}(t_\times)^2 - m_{ss}(t_\times)^2 
   +
    \left(\frac{d m_{aa}^2}{dt}\biggr\rvert_{t_\times}-\frac{d m_{ss}^2}{dt}\biggr\rvert_{t_\times}\right)(t-t_\times)
    \\
    \label{eq:LZTaylorExpand}
    &=  0 + \Delta' z \,.
\end{align}
In Eq.\,\eqref{eq:LZTaylorExpand}, we have simplified the expression $m_{aa}^2 - m_{ss}^2$ by making use of the fact that $  m_{aa}(t_\times)^2 - m_{ss}(t_\times)^2  = 0$, and defined $\Delta' = \frac{d}{dt}(m_{aa}^2 - m_{ss}^2)\rvert_{t_\times}$ and $z \equiv t - t_\times$ as a complexified time.

Inserting Eq.\eqref{eq:LZTaylorExpand} into Eq.\,\eqref{eq:massSqDifference} indicates that there is a level-crossing ($m_H^2 = m_L^2$), or turning-point in the integrand of Eq.\eqref{eq:adiabaticSolution}, at the complex times
\begin{align}
    \text{Complex Turning Points at }\pm z_0 = \pm 2 i \left|\frac{m_{as}^2}{\Delta'}\right| \, .
\end{align} 
The two turning point values at $\pm z_0$ are branch points associated with the root behavior of $m_H^2- m_L^2$, as shown by the blue ragged line of Fig.\,\ref{fig:LZComplexPlane}. Consequently, when analytically continuing the time integration variable $t$ in the argument of Eq.\,\eqref{eq:adiabaticSolution} to the complex time $z$, complex phases associated with wrapping around the branch occur, as given by
\begin{align}
    m_H - m_L &= \frac{m_H^2 - m_L^2}{m_H + m_L} 
    \nonumber \\
    &= \frac{\sqrt{(m_{aa}^2 - m_{ss}^2)^2 + 4 (m_{as}^2)^2}}{m_H + m_L} 
    \nonumber \\
    &\simeq \frac{\sqrt{(2m_{as}^2 + i\Delta' z)} \sqrt{(2m_{as}^2 - i\Delta' z)}}{m_H + m_L} 
     \nonumber
    \\
    &= \frac{\sqrt{(2 m_{as}^2)^2}}{m_H + m_L} \sqrt{z - z_0} \sqrt{z + z_0} = \frac{|2 m_{as}^2|}{m_H + m_L}|z -z_0|^{1/2}|z+z_0|^{1/2} e^{i \phi_1/2} e^{i \phi_2/2} e^{i 2\pi k/2} 
    \nonumber \\
    &= |m_H - m_L| e^{\frac{\scriptstyle i (\phi_1 + \phi_2 + 2\pi k)}{ \scriptstyle 2}}  \qquad \qquad \left(-\frac{\pi}{2} < \phi_i \leq \frac{3\pi}{2} , \; k \in \mathbb{Z} \right) \, .
    \label{eq:mHMinusmLPhases}
\end{align}
Here, $\phi_1$ ($\phi_2$) is the angle between the real axis and a ray extending from $+z_0$ ($-z_0$) to an arbitrary complex time $z$ as shown in Fig.\,\ref{fig:LZComplexPlane} \cite{brown2009complex}. The branch along the imaginary axis limits $\phi_i$ to lie between $-\pi/2$ and $3\pi/2$. The values of $\phi_1$, $\phi_2$, and $\exp[i(\phi_1 + \phi_2)/2]$ along the negative real axis (contour $\mathcal{C}_1$), positive real axis ($\mathcal{C}_4$), and on opposite sides of the branch ($\mathcal{C}_2$ and $\mathcal{C}_3$), are shown in the right panel of Fig.\,\ref{fig:LZComplexPlane}. The integer $k$ is arbitrarily chosen to be unity so that $m_H - m_L = +|m_H - m_L|$ on the negative real time axis.
\noindent
\begin{minipage}[t]{0.53\textwidth}
 \vspace{0pt}
    \centering
    \includegraphics[width=1.0\linewidth]{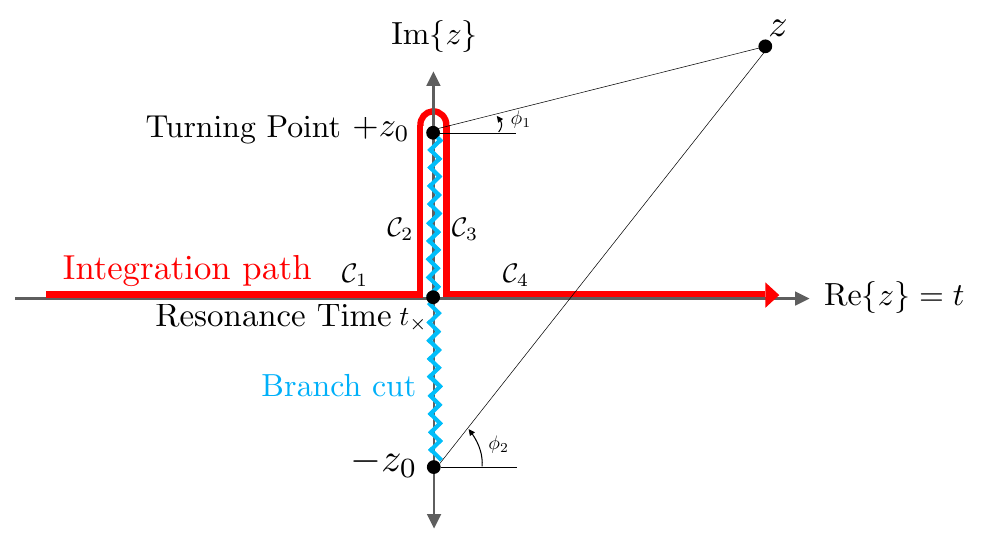} 
    \captionof{figure}{Analytic continuation of adiabatic evolution.}
    \label{fig:LZComplexPlane}
\end{minipage}
\hfill
\begin{minipage}[t]{0.44\textwidth}
 \vspace{0pt}
    \centering
    \setlength{\tabcolsep}{2pt}
    \renewcommand{\arraystretch}{2.575}
    \small{
    \begin{tabular}{l | c c c}
         & $\phi_1$ & $\phi_2$ & $\exp\left[i\frac{\phi_1 + \phi_2}{2}\right]$ \\
        \hline
        $\mathcal{C}_1$ \;& $\pi + \mathrm{arg}(z - z_0)$ & $\pi - \mathrm{arg}(z - z_0)$ & $-1$ \\
        $\mathcal{C}_2$ \;& $\frac{3\pi}{2}$ & $\frac{\pi}{2}$ & $-1$ \\
        $\mathcal{C}_3$ \;& $-\frac{\pi}{2}$ & $\frac{\pi}{2}$ & $1$ \\
        $\mathcal{C}_4$ \;& $-\mathrm{arg}(z - z_0)$ & $\mathrm{arg}(z - z_0)$ & $1$
    \end{tabular}}
    \label{tab:complexPhases}
    \captionof{table}{Phases around the branch.}
\end{minipage}

As can be seen from the Table \ref{tab:complexPhases} and Eq.\,\eqref{eq:mHMinusmLPhases}, $m_H - m_L$ flips sign as one integrates over the branch, which is equivalent to switching the energy evolution phase of $a_H$ with $a_L$ (see Eq.\,\eqref{eq:adiabaticSolution}). The connection goes even deeper. Indeed, recall the value of $\xi$ is inferred from the inverse of Eq.\,\eqref{eq:tan2xi}; $\xi(t) = \arctan(2 m_{as}^2/(m_{aa}^2 - m_{ss}^2))$. In the complex plane, $\arctan$ also has branch points precisely at $\pm z_0$, so that traversing around the branch makes $\xi$ discontinuous by $\xi \rightarrow \xi + \pi/2$ \cite{brown2009complex,arctanComplex}. While $\tan(2\xi +\pi) = \tan 2\xi$ remains the same, $\cos(\xi+\pi/2)=-\sin\xi$ and $\sin(\xi+\pi/2) =\cos(\xi)$, which is precisely the transformation that flips the relationship between $a_H$ and $a_L$ according to the relationship of Eq.\,\eqref{eq:rot}. Thus, $a_H$ and $a_L$ flip when passing to a new  branch.

Quantitatively, the trick of Landau is to perform the time integration of Eq.\,\eqref{eq:adiabaticSolution} past the resonance by integrating along the red contour of Fig.\,\ref{fig:LZComplexPlane}, which is always adiabatic since it avoids the non-adiabatic region at the time $t_\times + 0i$  \cite{Landau:1932vnv,landau1977quantum}. This integral can be split up into four segments, corresponding to the contours labeled $\mathcal{C}_1,\mathcal{C}_2,\mathcal{C}_3,\mathcal{C}_4$ of Fig.\,\ref{fig:LZComplexPlane}. For example, integrating Eq.\,\eqref{eq:adiabaticSolution} for $\tilde{a}_{H+}$ gives:
\begin{widetext}
\begin{align}
    \label{eq:LZContourIntegration}
    \tilde{a}_{H+}(t > t_\times) &=  \tilde{a}_{H+}(t_0)\sqrt{\frac{m_H(t_0)}{m_L(t)}}\exp\left[-i\int_{t_0 - t_\times}^{- \epsilon} \frac{\Delta}{2} dz -i\int_{-\epsilon}^{z_0-\epsilon} \frac{\Delta}{2}dz + i\int_{z_0+\epsilon}^{\epsilon} \frac{\Delta}{2}dz +i\int_{\epsilon}^{t} \frac{\Delta}{2}dz \right] \, ,
\end{align}
\end{widetext}
where $\Delta \equiv |m_H - m_L|$ for shorthand convenience.

Note the first two terms and last two terms have opposite signs, arising from being on opposite sides of the branch cut. In particular, the first term is a phase representing adiabatic evolution until right before $t_{\times}$, the second and third terms are equivalent, and the fourth term is the $\tilde{a}_{L+}(t)$ wavefunction, up to a phase. That is, wrapping around the branch converts an $\tilde{a}_H$ mass eigenstate into an $\tilde{a}_L$ mass eigenstate. The $\epsilon$ in the integration bounds makes explicit which side of the branch cut the contour resides, but is taken to $0$ at the end of the calculation. Similarly, the WKB coefficient $\sqrt{m_H(t_0)/m_H(t)}$ becomes  $\sqrt{m_H(t_0)/m_L(t)}$ after crossing the branch which follows from writing $m_H = \frac{1}{2}(m_H+m_L) + \frac{1}{2}(m_H-m_L)$ which becomes $\frac{1}{2}(m_H+m_L) - \frac{1}{2}(m_H-m_L) = m_L$ after the branch. Eq.\,\eqref{eq:LZContourIntegration} can therefore be written as
\begin{align}
    \label{eq:a1Toa2Wavefunction}
    \tilde{a}_{H+}(t > t_\times) = \tilde{a}_{H+}(t_0) \sqrt{\frac{m_H(t_0)}{m_L(t_0)}} \exp[-i \Phi] \exp\left[2i \int_0^{z_0} \frac{|m_H - m_L|}{2} dz\right] \frac{\tilde{a}_{L+}(t)}{\tilde{a}_{L+}(t_0)} \, ,
\end{align}
where $i\Phi = i\int_{t_0 - t_\times}^{-\epsilon} |m_H - m_L|$ is an unimportant complex phase, and $\tilde{a}_{L+}$ is the positive mode of the light energy eigenstate, given in Eq.\,\eqref{eq:adiabaticSolution}.
The residual integral from $0$ to $z_0$ along the branch, however, is entirely real and hence does not vanish when considering probabilities. The integral is
\begin{align}
    I &\equiv 2i\int_0^{z_0} \frac{m_H - m_L}{2} dz 
    \\
    &=\int_0^{z_0} \frac{i(m_H^2 - m_L^2)}{m_H + m_L} dz
    \\
    &\simeq\int_0^{z_0} \frac{i\sqrt{(\Delta')^2 z^2 + 4(m_{as}^2)^2}}{m_H(t_\times) + m_L(t_\times)} dz
    \\
    \label{eq:LZInt}
    &= -\left|\pi \frac{m_{as}^2}{\Delta'}\frac{1}{m_H(t_\times) + m_L(t_\times)} \right| \, .
\end{align}
Since $m_H$ and $m_L$ are nearly degenerate in the above integral, in the third line we approximate $m_H + m_L$ as a constant -- the value of their sum at the resonance time $t_\times$. 
\footnote{The integral can still be done analytically without this approximation, but it is complex and unnecessary in this work.} 

The same method, except by integrating below the branch cut instead of above, yields the  $\tilde{a}_{H-}$ solution after $t_\times$; it is the complex conjugate of Eq.\,\eqref{eq:a1Toa2Wavefunction}, as it must be, with $+$ fields replaced by their $ -$ counterparts. The linear combination of the $\tilde{a}_{H \pm}$ modes after $t_\times$ is then 
\begin{align}
    \label{eq:combinedaHLZ}
    \tilde{a}_H(t > t_\times) = \tilde{a}_{H+}(t>t_\times) +\tilde{a}_{H-}(t>t_\times)= \frac{\tilde{a}_H(t_0)}{\tilde{a}_L(t_0)} ) \sqrt{\frac{m_H(t_0)}{m_L(t_0)}} e^I [e^{i\Phi} \tilde{a}_{L+}(t) + e^{-i\Phi} \tilde{a}_{L-}(t)] \, .
\end{align}
After restoring the fields to their original definitions by the inverse phase factor $\tilde a_{\pm} \rightarrow \tilde{a}_{\pm} \exp[\mp i \int_{t_0}^t (m_H + m_L)/2]$, and inserting the definitions of $\tilde{a}_{L+}$ and $\tilde{a}_{L-}$, Eq.\,\eqref{eq:combinedaHLZ} reduces to
\begin{align}
    \label{eq:combinedaHLZ2}
    \tilde{a}_H(t>t_\times) = \tilde{a}_H(t_0)\sqrt{\frac{m_H(t_0)}{m_L(t)}}e^I \cos\left(\int_{t_0}^t m_L  d\tau + \Phi\right) \, ,
\end{align}
where we have used $|\tilde{a}_{L+}(t_0)| = |\tilde{a}_{L-}(t_0)|=\frac{1}{2}\tilde{a}_L(t_0)$.
Note, the right-hand-side of Eq.\,\eqref{eq:combinedaHLZ2} is an $\tilde{a}_L$ state. Hence the number density of $\tilde{a}_L$ after $t_\times$ that originate from $\tilde{a}_H$ before $t_\times$ is
\begin{align}
    n_{H\rightarrow L} &= \frac{1}{2m_L}\left[m_L^2 \tilde{a}_H(t>t_\times)^2 + (\partial_t\tilde{a}_H(t>t_\times))^2\right] = \frac{1}{2}\tilde{a}_H(t_0)^2 m_H(t_0) e^{2I}
    \\
    &= n_H^{\rm WKB} P_{\rm LZ}
\end{align}
where

\begin{align}
    \label{eq:LZProbabilityAppendix}
    &P_{\rm LZ} \equiv \exp[2I] \equiv \exp\left[-\frac{\pi \gamma}{2}\right] 
    \qquad \gamma =  \left|\frac{4(m_{as}^2)^2}{\frac{d{m_{aa}^2}}{dt} - \frac{d{m_{ss}^2}}{dt}} \frac{1}{(m_H + m_L)} \right|_{t = t_\times} \, ,
\end{align}
is the Landau-Zener probability of conversion, and is the main result of this appendix.

Note that the number density of $n_L$ after $t_\times$ also contains unconverted axions that originate from $\tilde{a}_L$ before $t_\times$ as well. In general, a wave function that starts off as $\tilde{a}_H$ before $t_\times$ will be after $t_\times$ a linear combination of $\tilde{a}_H$ and $\tilde{a}_L$, which are complete,
\begin{align}
    \tilde{a}(t > t_\times) = C_1(t) \tilde{a}_H(t) + C_2(t) \tilde{a}_L(t) \, .
\end{align}
As we have just demonstrated, $C_2(t)$, the projection of the new axion wave function in the $\tilde{a}_L$ direction, has a squared modulus given by Eq.\,\eqref{eq:LZProbabilityAppendix}. The coefficient $C_1(t)$ was lost along the integration of the contour in the upper complex plane because $\tilde{a}_H$ is exponentially suppressed relative to $\tilde{a}_L$ for positive imaginary time, as seen from Eq.\,\eqref{eq:adiabaticSolution}. Conversely, integrating Eq.\,\eqref{eq:adiabaticSolution} along a contour in the negative complex plane keeps only the $\tilde{a_H}$ projection, so that $C_1(t)$ can be extracted. It is simpler though to use conservation of probability over the brief duration of the resonance $\delta t_{\rm res}$, Eq.\,\eqref{eq:probabilityConservation}, to infer that $|C_1|^2 \simeq 1 - |C_2|^2$.

\subsection{Zener Approach}
\label{app:Zener_Approach}
A complementary approach, inspired by the work of Zener, solves the Schrödinger-like Eqns.\,\eqref{eq:FVHamiltonianApproxPlus} and \eqref{eq:FVHamiltonianApproxMinus} asymptotically via perturbation theory for a two-level system initially prepared in a state $\left|a_S\right>$ that due to the off-diagonal interaction evolves to $\left|a\right>$ at later times. 
To this aim, it is convenient to return to the flavor basis. Writing 
$(a_H \; a_L)^T = U^\dagger (a \; a_S)$, Eq.\,\eqref{eq:massEigenstates}, the Schrödinger equation of motion, Eq.\,\eqref{eq:schrodingerEqFeshbachVillars2}, becomes 
\begin{align}
    \label{eq:schrodingerEqFlavorBasis}
    i \partial_t
    \begin{pmatrix}
        \tilde{a}_{\pm}
        \\
        \tilde{a}_{S \pm}
    \end{pmatrix}
    &= 
    \pm
      \frac{m_H + m_L}{2}
    \begin{pmatrix}
        1 && 0
        \\
        0 &&  1
    \end{pmatrix}
    \begin{pmatrix}
        \tilde{a}_{\pm}
        \\
        \tilde{a}_{S \pm}
    \end{pmatrix}
    \pm \frac{m_H - m_L}{2}
    \begin{pmatrix}
        \cos(2\xi) &&  \sin(2\xi)
        \\
       \sin(2\xi) &&  -\cos(2\xi)
    \end{pmatrix}
    \begin{pmatrix}
        \tilde{a}_{\pm}
        \\
        \tilde{a}_{S \pm}
    \end{pmatrix} \, .
\end{align} 
Here, we have dropped the slow WKB evolution terms, $\dot{m_{H,L}}/m_{H,L}$, which are assumed to slowly change across the resonance.
Using the mixing angle definition, Eq.\,\eqref{eq:tan2xi}, the non-diagonal contribution of the Hamiltonian can be written as
\begin{align}
    \frac{m_H - m_L}{2}
    \begin{pmatrix}
        \cos(2\xi) &&  \sin(2\xi)
        \\
       \sin(2\xi) &&  -\cos(2\xi)
    \end{pmatrix} = 
    \frac{1}{2(m_H + m_L)}
    \begin{pmatrix}
        m_{aa}^2 - m_{ss}^2 &&  2 m_{as}^2
        \\
       2 m_{as}^2 &&  -(m_{aa}^2 - m_{ss}^2)
    \end{pmatrix}
    \equiv 
    \begin{pmatrix}
    \epsilon_1 (t) & \epsilon_{12} (t)\\
    \epsilon_{12} (t) & \epsilon_2 (t)
    \end{pmatrix}\,.
\end{align}
Defining $\Delta \epsilon \equiv \epsilon_1 - \epsilon_2 = (m_{aa}^2 - m_{ss}^2)/(m_H + m_L)$, the solutions to Eq.~\eqref{eq:schrodingerEqFlavorBasis} then follow:
\begin{alignat}{3}\label{eq:diff1}
    i \partial_t a_+ & = \epsilon_{12} \, a_{S+} \, e^{i \int_{t_0}^t \Delta \epsilon \, dt'}  \qquad i \partial_t a_- &= & -\epsilon_{12} \, a_{S-} \, e^{-i \int_{t_0}^t \Delta \epsilon \, dt'} \, \,, \\
    \label{eq:diff2}
    i \partial_t a_{S+} & = \epsilon_{12}^{*} \, a_+\, e^{- i\int_{t_i}^t \Delta \epsilon \, dt'} \qquad i \partial_t a_{S-} &=& -\epsilon_{12} \, a_-\, e^{ i\int_{t_i}^t \Delta \epsilon \, dt'} \,.
\end{alignat}
As expected, the positive and negative frequency solutions are Hermitian conjugates of each other. For brevity, we will therefore focus on the just the positive solution modes with understanding that the negative frequency modes are the complex conjugated solutions. The initial conditions of each axion state are
\begin{align}
    a_+ (t_0) &= a_- (t_0)^* = \frac{1}{2}a(t_0) =0\,,\\
    a_{S+} (t_0) &=a_{S-}^* (t_0) = \frac{1}{2}a_{S} (t_0)\,.
\end{align}
Equations~\ref{eq:diff1} and \ref{eq:diff2}
can be combined into a second-order differential equation that decouples the two states. For example, for the positive mode,
\begin{equation}
\label{eq:Lz_num}
    \ddot a_+ - \left(i \Delta \epsilon + \frac{\dot{ \epsilon_{12}}}{\epsilon_{12}}\right) \dot a_+ + \epsilon_{12}^2 a_+ = 0\,.
\end{equation}

Assuming that $\epsilon_{12}$ is constant over the resonance, and that Taylor expanding $\Delta \epsilon$ about the resonance is linear in time (see Eq.\,\eqref{eq:LZTaylorExpand}), this equation can be solved analytically in terms of parabolic cylinder functions \cite{abramowitz1968handbook}; see~\cite{Carenza:2023nck, Ivakhnenko:2022sfl,HammarskioldSpendrup1879424} for recent reviews. The solution is \cite{Carenza:2023nck}
\begin{align}
    a_+(t >t_\times) = a_{S+}(t_0)\sqrt{\frac{\gamma}{4}} e^{-\pi \gamma}D_{-1-n}\left(\frac{iw(t)}{\sqrt{n}}\right)\, ,
\end{align}
where $\gamma$ is the adiabatic parameter given in Eq.\,\eqref{eq:LZProbabilityAppendix}, $n = i \gamma/4$, $w(t)= \sqrt{\Delta \epsilon (t-t_\times)}$, and $D_\nu(z)$ is the parabolic cylindrical function \cite{abramowitz1968handbook}.
In particular, the probability that the system, initially prepared in the sterile flavor state, is identified with the active axion flavor state after crossing is
\begin{equation}\label{eq:PZener}
    \lim_{t\to \infty} \left|\frac{a_+(t>t_\times)}{a_{S+}(t_0)}\right|^2 = 1- \exp[-\pi \gamma/2]\,,\quad \text{with}\quad \gamma = \left|4\epsilon_{12}^2\left(\frac{d \Delta \epsilon}{dt}\right)^{-1}\right|_{t=t_\times} =  \left|\frac{4(m_{as}^2)^2}{\frac{d{m_{aa}^2}}{dt} - \frac{d{m_{ss}^2}}{dt}} \frac{1}{(m_H + m_L)} \right|_{t = t_\times}\,,
\end{equation}
which by the same logic as in Appendix \ref{app:Landau_Approach}, is the same probability of converting $a \rightarrow a_S$.

At first sight, this looks similar to the conversion probability derived in Eq.\,\eqref{eq:LZProbabilityAppendix}, except the probability here is $1-P_{\rm LZ}$ rather than $P_{\rm LZ}$. The difference lies in that after the resonance at $t_\times$, the main flavor state identified with the mass  eigenstates flips. That is \cite{kuo1989nonadiabatic},
\begin{align}
\begin{split}
    P_{a_S \rightarrow a} &= |\langle a(t>t_\times)| a_S(t_0)\rangle|^2
    \\
    &= |\langle a(t>t_\times)|a_i(t>t_\times)\rangle \langle a_i(t>t_\times| a_j(t_0)\rangle\langle a_j(t_0)| a_S(t_0)\rangle|^2
    \\
    &= 
    \begin{pmatrix}
        0 & 1
    \end{pmatrix}
    \begin{pmatrix}
        \cos^2\xi(t>t_\times) & \sin^2\xi(t>t_\times)
        \\
         \sin^2\xi(t>t_\times) & \cos^2\xi(t>t_\times)
    \end{pmatrix}
    \begin{pmatrix}
        1-P_{\rm LZ} & P_{\rm LZ}
        \\
         P_{\rm LZ} &  1-P_{\rm LZ}
    \end{pmatrix}
    \begin{pmatrix}
        \cos^2\xi(t_0) & \sin^2\xi(t_0)
        \\
         \sin^2\xi(t_0) & \cos^2\xi(t_0)
    \end{pmatrix}
    \begin{pmatrix}
        1 & 0
    \end{pmatrix}
    \\
    \label{eq:probAtoAs}
    &= \frac{1}{2}\left[1+ (2 P_{\rm LZ} -1) \cos( 2 \xi(t>t_\times) \cos (2 \xi(t_0)  \right] \, .
\end{split}
\end{align}
Here, we have inserted the resolution of the identity as a sum of outer products of the axion mass eigenstates in line 2, and used the unitary operator that rotates flavor eigenstates to mass eigenstates in line 3. To map on to the Landau approach of App.\ref{app:Landau_Approach}, we define as usual $P_{\rm LZ} \equiv |\langle a_i(t>t_\times)|a_j(t_0)\rangle|^2$ for $i \neq j$ as the probability of converting mass eigenstate $i$ to mass eigenstate $j$.

Since prior to $t_\times$, $m_{aa}^2 -m_{ss}^2 \ll m_{as}^2$ (or vice versa) and after $t_\times$, $m_{aa}^2 - m_{ss}^2 \gg m_{as}^2$ (or vice versa), $2\xi \simeq \pi$ before $t_\times$ and $2\xi \simeq 0$ after $t_\times$ (or vice versa), as can be seen from Eq.\eqref{eq:tan2xi}. Hence the product $\cos( 2 \xi(t>t_\times) \cos (2 \xi(t_0) \simeq -1$ and Eq.\eqref{eq:probAtoAs} can be approximately written as
\begin{align}
    \label{eq:}
    P_{a_S\rightarrow a} \simeq 1 - P_{\rm LZ} \, .
\end{align}

Comparing this with Eq.\,\eqref{eq:PZener}, we can identify $P_{\rm LZ} = \exp[-\pi \gamma/2]$, which is consistent with the Landau approach of App.\ref{app:Landau_Approach}.

\subsection{Validity of Landau-Zener Probability}
\label{app:LZValidity}

Importantly, the analytic integration of Eq.~\ref{eq:LZInt} in the Landau approach or Eq.~\ref{eq:Lz_num} in the Zener approach used to derive the LZ probability are obtained under two important assumptions: \textit{(1)} $m_{aa}^2 - m_{ss}^2$ is linear with time about $t_\times$ and \textit{(2)} the variation of the off-diagonal term, $m_{as}^2$, is negligible in comparison to that of $m_{aa}^2 - m_{ss}^2$ during the resonance.

To check the first condition, let us fully Taylor expand $m_{aa}^2 -m_{ss}^2$ about $t_{\times}$ to determine when higher order terms in the expansion dominate over the linear term:
\begin{align}
\label{eq:LZTaylorExpandLinearCheck}
    m_{aa}^2 -m_{ss}^2
    &= 0 + (t-t_{\times}) \frac{d \Delta}{dt} + \frac{1}{2}(t - t_{\times})^2\frac{d^2 \Delta}{dt^2} +  \frac{1}{6}(t - t_{\times})^3\frac{d^3\Delta}{dt^3} + ... \,.
\end{align}
Here, $\frac{d^k}{dt^k}\Delta \equiv \frac{d^k}{dt^k}(m_{aa}^2 - m_{ss}^2)\biggr\rvert_{t_\times}$  as before, which is bounded by
\begin{align}
    \frac{d^k}{dt^k} \Delta &\leq \frac{d^k}{dt^k}m^2 \biggr\rvert_{t_\times}
    \\
    &\leq \frac{d^k}{dt^k} \left(m_{a,0} \left(\frac{T}{T_{\rm QCD}}\right)^{-n} \right)^2 \biggr\rvert_{t_\times}
    \\
    &=\frac{d^{k-1}}{dt^{k-1}} \left(2n H m^2 \right)\biggr\rvert_{t_\times}
    \\
    \label{eq:LZderivativeOrder}
    &= (2n H(t_\times))^k  m(t_\times)^2 \, .
\end{align}
In the third line, we have used entropy conservation to write $dT/dt = -H T$, where $H$ is Hubble. In the fourth line, we have dropped terms of order $(dH/dt)^l H^{k-l}$ which are small compared to terms of order $H^k$ in a radiation or matter-dominated cosmology.

From Eq.\,\eqref{eq:LZderivativeOrder}, we see that each term in Eq.\,\eqref{eq:LZTaylorExpandLinearCheck} is smaller (or larger) than the previous term by a factor of at least $2 H(t_\times)(t - t_\times)$. Thus, if $2 H(t_\times)\delta t_{\rm res} < 1$, where $\delta t_{\rm res}$ is the duration of the resonance where LZ is active, then the linear condition assumed for LZ is valid.

Quantitatively, it is convenient to define $\delta t_{\rm res}$ as the full width at half max of the $\sin^2 2\theta$ peak  \cite{kuo1989nonadiabatic}, meaning if $\sin^2 2\theta(t_\times) = 1$, then $\sin^2 2\theta(t_{\times} + \delta t_{\rm res}) = 1/2$. From Eq.\,\eqref{eq:tres}, this defines 
\begin{align}
    \delta t_{\rm res} = \left|\frac{2 m_{as}^2(t_\times)}{\Delta'}\right| = 2
    \sqrt{\gamma\frac{m_1(t_\times) + m_2(t_\times)}{\Delta'}} \, ,
\end{align}
where we have used the definition of the adiabatic parameter, Eq.\,\eqref{eq:gamma}, to replace $m_{as}^2$ with $\gamma$. Unless there is a fine-tuned cancellation between $d m_{aa}^2/dt$ and $dm_{ss}^2/dt$ at $t_{\times}$, the typical value of $\Delta' \sim 2 n H(t_\times) m(t_\times)^2$, which is Eq.\,\eqref{eq:LZderivativeOrder} for $k = 1$. If we define $\delta$ as the degree of tuning of the two terms in $\Delta'$ compared to this expected value, ie $\Delta' \equiv 2 n H(t_{\times})m(t_\times)^2 \delta$ where $\delta \leq 1$, then linearity of LZ requires
\begin{align}
    \label{eq:LZlinearityCondition}
    \delta t_{\rm res} H(t_\times) \approx 2 \sqrt{\frac{\gamma}{n \delta}\frac{H(t_{\times})}{m(t_\times)}} \ll 1 \qquad \text{(Linearity Condition of LZ)}\, \,
\end{align}
which is satisfied unless $\delta \ll m(t_\times)/H(t_\times) n\gamma$, which is a small parameter since $m > H$ and $\gamma < 1$ in the non-adiabatic LZ regime. For example, for the axion mass matrix of Eq.\,\eqref{eq:massMatrix}, $\delta = 1-R_f^2$. Since level crossing does not occur for $R_f > 1$ for the potential of \eqref{eq:massMatrix}, $\delta$ is $\mathcal{O}(1)$ except potentially for some  $R_f$ very close to $1$. Similarly, for the axion mass matrix of Eq.\,\eqref{eq:VII}, $\delta = 1 - R_f^{-2}$, which agrees with the expected mapping of $R_f \rightarrow R_f^{-1}$ between the two example potentials as discussed in Sec.\,\ref{sec:PotentialII}. Since level crossing only occurs for $R_f > 1$ for this potential, $\delta$ is again $\mathcal{O}(1)$ for all of the parameter space except potentially near $R_f = 1$.

Last, to check the second condition, we note that $dm_{as}^2/dt \leq 2 n H m_{as}^2$. It follows that if the linearity condition of LZ is satisfied, \eqref{eq:LZlinearityCondition}, then the rate of change of $m_{aa}^2 - m_{ss}^2$ compared to the rate of change of $m_{as}^2$ around $t_\times$ is
\begin{align}
    \label{eq:ratioOfDiagonaltoOffDiagonal}
    \frac{\frac{d}{dt}(m_{aa}^2 -m_{ss}^2)}{\frac{d}{dt}(m_{as}^2)} &=\frac{\Delta'}{\frac{d}{dt}(m_{as}^2)} 
    \\
    &\geq \frac{2n H(t_\times) m(t_\times)^2 \delta}{2n H(t_\times) m_{as}^2(t_\times)}
    \\
    & \geq \frac{2 m(t_\times)^3 /\gamma}{2n H(t_\times) m_{as}^2(t_\times)}
    \\
    & \geq \frac{m(t_\times)^2}{ m_{as}^2(t_\times)}\frac{1}{n \gamma}
\end{align}
Thus, the range of change of $m_{as}^2$ can always be ignored in the LZ formalism if $m_{as}^2(t_\times)  < m^2/n\gamma$, where $m^2 = m_{aa}^2(t_\times) = m_{ss}^2(t_\times)$.
Since $\gamma \leq 1$ in the non-adiabatic regime, violation of this condition requires the off-diagonal element $m_{as}^2$, to be larger than either of the diagonal elements, $m_{aa}^2$ or $m_{ss}^2$, around $t_\times$, which is often not the case. For example, for the axion mass matrix of Eq.\,\eqref{eq:massMatrix}, the ratio of Eq.\,\eqref{eq:ratioOfDiagonaltoOffDiagonal} is infinite since $m_{as}^2 = m_S^2 R_f$ is independent of time. Hence ignoring the rate of change of $m_{as}^2$ is always satisfied. Similarly, for the axion mass matrix of Eq.\,\eqref{eq:VII}, the ratio of Eq.\,\eqref{eq:ratioOfDiagonaltoOffDiagonal} is $R_f(1-R_f^{-2})$. This ratio is satisfied for all $R_f$ except for $R_f \sim 1$ where the LZ linearity condition anyway breaks down.

\newpage
\bibliography{lib}

\end{document}